\documentclass[aps,prd,amsmath,twocolumn,preprintnumbers]{revtex4}
\usepackage{graphicx}

\begin{document}

\date{\today}

\title{Boson stars and oscillatons in an inflationary universe}

\author{Gyula Fodor$^1$, P\'eter Forg\'acs$^{1,2}$,
 M\'ark Mezei$^{3}$}
\affiliation{$^1$MTA RMKI, H-1525 Budapest 114, P.O.Box 49, Hungary,\\
$^2$LMPT, CNRS-UMR 6083, Universit\'e de Tours, Parc de Grandmont,
37200 Tours, France \\
$^3$Center for Theoretical Physics, Massachusetts Institute of Technology,
Cambridge, Massachusetts 02139, USA
}

\begin{abstract}
Spherically symmetric gravitationally bound, oscillating scalar lumps
(boson stars and oscillatons) are considered in Einstein's gravity
coupled to massive scalar fields in $1+D$ dimensional
de Sitter-type inflationary space-times.
We show that due to inflation bosons stars and oscillatons
lose mass through scalar radiation, but at a rate
that is exponentially small when the expansion rate is slow.
\end{abstract}

\preprint{MIT-CTP 6-4161}

\maketitle

\section{Introduction}

Current physical field theories explaining the presently known
fundamental interactions are remarkably successful to account for
phenomena both on particle physics (i.e.\ $10^{-18}$m ) and on
cosmological ($10^{26}$m) scales, usually incorporate various
fundamental scalar fields (Higgs bosons, axion, supersymmetric
partners of fermions, inflaton, waterfall fields etc.).  It has been
observed since a long time ago that self-interacting scalar fields
admit at least up to $D=4$ spatial dimensions spatially localized,
oscillating lumps, nowadays called
oscillons\cite{Dashen,BogMak2,CopelGM95}. Oscillons are reminiscent
of spatially localized, time-periodic breathers known in one
dimensional Sine-Gordon theory.  There is, however, a fundamental
difference, in that oscillons are only {\sl approximatively}
time-periodic since they radiate, thereby losing continuously their
energy, albeit very slowly \cite{SK,FFHM,moredim}.  Moreover
oscillon-type objects exist in a broad class of field theories
containing in addition to scalar, vector and possibly other fields
\cite{GleisThor,GleiserTHor09}. A prominent theory admitting classical
oscillons is the bosonic sector of the Standard Model
\cite{Farhi05,Graham07a,Graham07b}.  Einstein's gravity coupled to a
free, massive real Klein-Gordon field, also possesses oscillon-type
solutions, which go under the name of oscillatons
\cite{Seidel1,Seidel2,LopezMatos,Alcub,Page,Balak}.

An important problem is to investigate the possible pertinence of
oscillatons for cosmology where according to current theories some
scalars fields are necessarily involved.  In recent investigations the
evolution of oscillons in an inflating background has been
investigated
\cite{Broadhead:2005hn,gleiserijmp,GrahStam,fggirs,amin,amin2}.  It
has been found that an inflating background also induces a radiative
tail in oscillons, {\sl distinct} from the one present in flat
space-times, leading to an additional leakage of energy.  However this
decay rate is also exponentially suppressed, so inflating oscillons
may survive for an exponentially long time provided that the horizon
is far larger than the width of an oscillon.  As recently shown, due
to quantum effects, the decay rate of oscillons becomes power law-like
and in many (but not in all) cases dominates over the classical decay
rate \cite{hertzberg}.  There are indications that oscillons can be
copiously produced from thermal initial conditions and a substantial
fraction of the energy of the system will be stored by them
\cite{Broadhead:2005hn,gleisergraham,amin2}. Therefore it is important to
investigate the physical r\^ole of such long living localized objects.  In a
cosmological setting it is necessary to take into account the
influence of gravity, i.e.\ instead of simply considering oscillons in
a background metric, to investigate solutions of the {\sl coupled
  Einstein-scalar system} -- {\sl oscillatons} -- with an expanding,
asymptotically homogeneous metric.

In this paper we undertake a thorough investigation of localized,
oscillaton-type solutions in Einstein's gravity coupled to a massive
Klein-Gordon field, when space-time undergoes inflation, driven by a
negative pressure cosmological fluid -- i.e.\ including a cosmological
constant, $\Lambda>0$. (For recent reviews on inflation theory see
e.g.~\cite{LythLiddle,weinberg,mukhanov}.)

Our results show that once admitting massive scalar fields, both in
present day cosmology and in the early Universe, spatially localized,
very long living oscillatons exist.  The inflaton field generates an
energy density of the order of $(10^{16}GeV)^4$ and negative pressure
which can be described by an effective (and slowly varying)
cosmological parameter. Using the standard slow roll approximation for
the inflaton, the metric during the inflationary era can be
approximated very well with a de Sitter one. The order of magnitude of
the cosmological constant during the inflationary epoch is
$\Lambda\approx10^{-11}$ (in Planck units). If other massive scalar
fields are present during inflation such as the waterfall-field in
hybrid inflationary scenarios \cite{linde1,linde2}, oscillaton
production is expected to be significant. We find that oscillatons
may survive on cosmological time-scales only if the mass of the scalar
field is larger than $10^{15}GeV/c^2$. The mass loss of an oscillaton
during Hubble time is completely determined by the ratio
$\mu_c/\mu_\Lambda$, where $\mu_\Lambda$ is the energy density due to
$\Lambda$ and $\mu_c$ is that of the oscillaton in its center.  For
example when $\mu_c/\mu_\Lambda\approx 55$ the mass loss of such a
``waterfall'' oscillaton is of the order of $1\%$ during Hubble time.
This mass loss is exponentially decreasing for increasing values of
the ratio $\mu_c/\mu_\Lambda$.

Another interesting scenario for oscillaton production is during
reheating, assuming that the inflaton field undergoes oscillations
near the minimum of its potential.  These oscillatons store a
considerable amount of energy and due to parametric resonances induced
by quantum effects (for oscillons this has been shown in Ref.\
\cite{hertzberg}) they may influence the fluctuation spectrum.

The present value of $\Lambda\approx10^{-122}$ (in Planck units) being
extremely small, its effect on the mass loss of generic oscillatons is
completely negligible on a cosmological time scale.  For example if
the mass of the scalar field is of the order of $m\approx10^{-15}$eV
and the amplitude of the oscillaton is approximatively $10^{-17}$, the
effect of the cosmological constant results in a roughly $1\%$ mass
loss during the lifetime of the Universe.

We also consider (spherically symmetric) boson stars
\cite{Kaup,Ruffini,Jetzer,SchunkMielke} in this setting,
i.e.\ when a complex scalar field is coupled to gravity with a nonzero
cosmological constant.  Adopting our small amplitude expansion
techniques developed for oscillons and oscillatons we obtain boson
star solutions in a systematic way.  Remarkably, this method yields
very good results even for near maximal mass boson stars.  Spatially
well localized boson star-type solutions with a static metric,
familiar in the case $\Lambda\leq0$ no longer exist when $\Lambda>0$.
There exist, however, radiating boson stars with a time dependent
metric loosing continually their mass due to emission of scalar waves.
We obtain their mass loss using our methods developed for oscillons
resp.\ oscillatons, with the result that this mass loss is
exponentially small and it is negligible for sufficiently small values
of $\Lambda$.

\section{Field equations}

We consider a real or complex scalar field coupled to gravity.  The
total Lagrangian density is
\begin{equation}
\mathcal{L}=\frac{1}{16\pi}\mathcal{L}_G+\mathcal{L}_M \,,
\label{acttot}
\end{equation}
where the Einstein Lagrangian density in case of a cosmological
constant $\Lambda$ is
\begin{equation}
\mathcal{L}_G=\sqrt{-g}(R-2\Lambda) \,.
\end{equation}
We work in a $D+1$ dimensional curved space-time with metric $g_{ab}$
and assume metric signature $(-+++...)$. Unless explicitly noted, we
use Planck units $G=c=\hbar=1$.
In some cases
it is advantageous to use natural units, where $c=\hbar=1$ (but
$G\not=1$) and all quantities are measured in various powers of
electron volts.  Because of the different notation in the potential
term we consider the real and the complex field case separately.
Variation of the action \eqref{acttot} with respect to $g^{ab}$ yields
Einstein equations,
\begin{equation}
G_{ab}+\Lambda g_{ab}=8\pi T_{ab} \,. \label{eq:einst}
\end{equation}

\subsection{Boson stars}

We consider a complex scalar field $\Phi$ with a self-interaction
potential $U(\Phi^*\Phi)$. For a free field with mass $m$ the
potential is $U(\Phi^*\Phi)=m^2\Phi^*\Phi$.  The Lagrangian density
belonging to the scalar field is
\begin{equation}
\mathcal{L}_M=-\frac{\sqrt{-g}}{2}\left[g^{ab}\Phi_{,a}^*\Phi_{,b}
+U(\Phi^*\Phi)\right] \,.
\end{equation}
Variation of the action \eqref{acttot} with respect to $\Phi^*$ yields
the wave equation
\begin{equation}
g^{ab}\Phi_{;ab}-\Phi\,U'(\Phi^*\Phi)=0 \,, \label{eq:wave1}
\end{equation}
where the prime denotes derivative with respect to $\Phi^*\Phi$.  The
stress-energy tensor in this case is
\begin{equation}
T_{ab}=\frac{1}{2}\left\{\Phi_{,a}^*\Phi_{,b}
+\Phi_{,b}^*\Phi_{,a}
-g_{ab}\left[g^{cd}\Phi_{,c}^*\Phi_{,d}+U(\Phi^*\Phi)\right]\right\}
. \label{eq:tab1}
\end{equation}

We shall assume that the self-interaction potential, $U(\Phi^*\Phi)$,
has a minimum $U(\Phi^*\Phi)=0$ at $\Phi=0$, and expand its derivative
as
\begin{equation}
U'(\Phi^*\Phi)=m^2\left[1+\sum\limits_{k=1}^{\infty}u_k(\Phi^*\Phi)^k
\right] ,
\end{equation}
where $m$ is the scalar field mass and $u_k$ are constants.

In order to get rid of the $8\pi$ factors in the equations we
introduce a rescaled scalar field and potential by
\begin{equation}
\phi=\sqrt{8\pi}\,\Phi \ , \qquad
\bar U(\phi^*\phi)=8\pi U(\Phi^*\Phi) \,.  \label{eq:rscpu1}
\end{equation}
Then
\begin{equation}
\bar U'(\phi^*\phi)=m^2\left[1+\sum\limits_{k=1}^{\infty}
\bar u_k(\phi^*\phi)^k\right],
\end{equation}
with
\begin{equation}
\bar u_k=\frac{u_k}{(8\pi)^{k}} \,.
\end{equation}

\subsection{Oscillatons}

In this case we consider a real scalar field $\Phi$ with a
self-interaction potential $U(\Phi)$. For a free field with mass $m$
the potential is $U(\Phi)=m^2\Phi^2/2$.  The Lagrangian density
belonging to the scalar field is
\begin{equation}
\mathcal{L}_M=-\sqrt{-g}\left(\frac{1}{2}\Phi_{,a}\Phi^{,a}
+U(\Phi)\right) \,.
\end{equation}
Variation of the action \eqref{acttot} with respect to $\Phi$ yields
the wave equation
\begin{equation}
g^{ab}\Phi_{;ab}-U'(\Phi)=0 \,, \label{eq:wave2}
\end{equation}
where the prime now denotes derivative with respect to $\Phi$.  The
stress-energy tensor is
\begin{equation}
T_{ab}=\Phi_{,a}\Phi_{,b}-g_{ab}
\left(\frac{1}{2}\Phi_{,c}\Phi^{,c}+U(\Phi)\right) . \label{eq:tab2}
\end{equation}

We shall assume that $U(\Phi)$ has a minimum $U(\Phi)=0$ at $\Phi=0$,
and expand its derivative as
\begin{equation}
U'(\Phi)=m^2\left(\Phi+\sum\limits_{k=2}^{\infty}u_k\Phi^k
\right) ,
\end{equation}
where $m$ is the scalar field mass and $u_k$ are constants.

In order to get rid of the $8\pi$ factors in the equations we
introduce a rescaled scalar field and potential by
\begin{equation}
\phi=\sqrt{8\pi}\,\Phi \ , \qquad
\bar U(\phi)=8\pi U(\Phi) \,.  \label{eq:rscpu2}
\end{equation}
Then
\begin{equation}
\bar U'(\phi)=m^2\left(\phi+\sum\limits_{k=2}^{\infty}
\bar u_k\phi^k\right) ,
\end{equation}
with
\begin{equation}
\bar u_k=\frac{u_k}{(8\pi)^{(k-1)/2}} \,.
\end{equation}

\subsection{Scaling properties}

In both the real and the complex case, if the pair $\phi(x^c)$ and
$g_{ab}(x^c)$ solves the field equations with a potential $\bar U$ and
cosmological constant $\Lambda$, then
\begin{equation}
\hat\phi(x^c)=\phi(\gamma x^c) \ , \qquad
\hat g_{ab}(x^c)=g_{ab}(\gamma x^c)   \label{scaleprop}
\end{equation}
for any positive constant $\gamma$, is a solution with a rescaled
potential $\gamma^2\bar U$ and rescaled cosmological constant
$\gamma^2\Lambda$. This scaling property may be used to set the scalar
field mass to any prescribed value, for example to make $m=1$.

If $D=1$ then, by definition, the Einstein tensor is traceless, and
from the trace of the Einstein equations it follows that the potential
$U$ is constant. Hence we assume that $D>1$.

\subsection{Spherically symmetric $D+1$ dimensional space-time}

We consider a spherically symmetric $D+1$ dimensional space-time using
isotropic coordinates $x^\mu=(t,r,\theta_1,...,\theta_{D-1})$. In this
coordinate system the metric is diagonal and its spatial part is in
conformally flat form, with components
\begin{equation}\label{eq:metrgen}
\begin{split}
g_{tt}&=-A \ , \qquad\, g_{rr}=B \ ,\\
g_{\theta_1\theta_1}&=r^2 B \ , \qquad
g_{\theta_n\theta_n}=r^2 B\prod_{k=1}^{n-1}\sin^2\theta_k \ ,
\end{split}
\end{equation}
where $A$ and $B$ are functions of temporal coordinate $t$ and radial
coordinate $r$. In case of oscillatons, the periodic change in the
acceleration of the constant radius observers is much smaller in
isotropic coordinates than in the more commonly used Schwarzschild
coordinate system \cite{Page,oscillaton}. It is also more convenient
to study the Newtonian limit of boson stars using isotropic
coordinates \cite{friedberg}. The components of the Einstein tensor
are
\begin{align}
G_{tt}&=\frac{D-1}{2}\Biggl[
\frac{D}{4B^2}\left(B_{,t}\right)^2 \notag\\
&\qquad\qquad-\frac{A}{r^{D-1}B^{\frac{D+2}{4}}}
\left(\frac{r^{D-1}B_{,r}}{B^{\frac{6-D}{4}}}\right)_{,r}
\Biggr] , \label{eq:gtt}\\
G_{rr}&=\frac{D-1}{2}\Biggl[
\frac{(D-2)\left(r^2B\right)_{,r}}{4r^4A^{\frac{2}{D-2}}B^2}
\left(r^2A^{\frac{2}{D-2}}B\right)_{,r} \notag\\
&\qquad-\frac{1}{A^{\frac{1}{2}}B^{\frac{D}{4}-1}}
\left(\frac{B^{\frac{D}{4}-1}B_{,t}}{A^{\frac{1}{2}}}
\right)_{,t}
-\frac{D-2}{r^2}
\Biggr] , \label{eq:grr}\\
G_{tr}&=-\frac{D-1}{2}\,A^{\frac{1}{2}}\left(\frac{B_{,t}}
{A^{\frac{1}{2}}B}\right)_{,r} , \label{eq:gtr}\\
G_{\theta_1\theta_1}&=r^2G_{rr}
+\frac{r^3B}{2A^{\frac{1}{2}}}\left(\frac{A_{,r}}
{rA^{\frac{1}{2}}B}\right)_{,r} \notag\\
&\qquad\qquad+\frac{D-2}{2}r^3B^{\frac{1}{2}}
\left(\frac{B_{,r}}{rB^{\frac{3}{2}}}\right)_{,r} . \label{eq:gthth}
\end{align}
The potential independent term in the wave equation \eqref{eq:wave1}
and \eqref{eq:wave2} is
\begin{align}
g^{ab}\phi_{;ab}&=\frac{\phi_{,rr}}{B}
-\frac{\phi_{,tt}}{A}
-\frac{\phi_{,t}}{2B^{D}}\left(\frac{B^{D}}{A}\right)_{,t}
\notag\\
&+\frac{\phi_{,r}}{2r^{2D-2}AB^{D-1}}\left(r^{2D-2}AB^{D-2}\right)_{,r}
 \,. \label{eq:waveterm}
\end{align}

\subsection{de Sitter space-time}

If the cosmological constant $\Lambda$ is positive, far from the
central boson star or oscillaton the space-time should approach the de
Sitter metric.  In static Schwarzschild coordinates on a $D+1$
dimensional space-time this metric can be written as
\begin{equation}
ds^2=-\left(1-H^2\bar r^2\right)dt^2
+\frac{d\bar r^2}{1-H^2\bar r^2}
+\bar r^2d\Omega^2_{D-1} \ ,
\end{equation}
where $d\Omega^2_{D-1}$ is the metric on the unit sphere $S^{D-1}$.
The Hubble constant $H$ is related to the cosmological constant
$\Lambda$ by
\begin{equation}
H^2=\frac{2\Lambda}{D(D-1)} \ . \label{caphlambda}
\end{equation}
In isotropic coordinates the de Sitter metric takes the form
\begin{equation}
ds^2=-\frac{\Bigl(1-\frac{H^2r^2}{4}\Bigr)^2}
{\Bigl(1+\frac{H^2r^2}{4}\Bigr)^2}\,dt^2
+\frac{dr^2+r^2d\Omega^2_{D-1}}
{\Bigl(1+\frac{H^2r^2}{4}\Bigr)^2} \ ,     \label{desittermet}
\end{equation}
where the two radial coordinates are related by
\begin{equation}
\bar r=\frac{r}{1+\frac{H^2r^2}{4}} \ .
\end{equation}
The cosmological horizon, which is at $\bar r=1/H$ in the
Schwarzschild coordinate system, is at $r_h=2/H$ in isotropic
coordinates.

\section{Small-amplitude expansion} \label{sec:smallampl}

The small-amplitude expansion procedure has been applied successfully
to describe the core region of flat background oscillons
\cite{Dashen,SK,Kichenassamy,FFHL}.  The method was generalized in
\cite{oscillaton} to the case when the scalar field is coupled to
Einstein gravity, and in \cite{dilaton} to a very similarly behaving
scalar-dilaton system. In this section we expand oscillatons in the
case when the cosmological constant is positive.

Newtonian boson stars have already been investigated in the paper of
Ruffini and S. Bonazzola \cite{Ruffini}, obtaining a system of two
coupled differential equations, which nowadays are generally called
Schr\"odinger-Newton equations in the
literature\cite{Diosi,Penrose,Moroz,Tod}. The same system was obtained
as weak gravity limit of general relativistic boson stars in
\cite{Friedberg} and \cite{Ferrell}. The leading order results of the
small-amplitude expansion of boson stars, that we present in this
section, when restricted for $\Lambda=0$, also yield the same
equations.

The repulsive effect of the cosmological constant is taken into
account in a similar way as in \cite{GrahStam,fggirs,amin}. Since we
intend to investigate localized objects, where at large distances the
scalar field is negligible, by the generalized Birkhoff's theorem
\cite{eiesland,morrow-witt,schleich}, the metric should approach the
Schwarzschild-de Sitter metric. This also means that in the asymptotic
de Sitter region the expansion rate (Hubble parameter) is constant.

\subsection{Choice of coordinates}

We are looking for spatially localized bounded solutions of the field
equations for which $\phi$ is small and the metric is close to de
Sitter. The smaller the amplitude of a boson star or an oscillaton is,
the larger its spatial extent becomes. Therefore, we introduce a new
radial coordinate $\rho$ by
\begin{equation}
\rho=\varepsilon m r \,,
\end{equation}
where $\varepsilon$ denotes the small-amplitude parameter. Motivated
by the scaling property \eqref{scaleprop} we have also included a $m$
factor into the definition of $\rho$.

Long-living localized solutions are expected to exist only if their
size is considerably smaller than the distance to the cosmological
horizon.  We introduce a rescaled Hubble constant $h$ by
\begin{equation}
H=\varepsilon^2 m h \,,  \label{caphlowh}
\end{equation}
and assume that even $h$ is reasonably small. This way we ensure that
the typical size of boson stars or oscillatons, which is of the order
$1/(\varepsilon m)$ using the $r$ coordinates, remains smaller than
the radius of the cosmological horizon, $r_h=2/H=2/(\varepsilon^2 m
h)$.

We will see in Section \ref{sec:enden} that the energy density $\mu$
of the scalar field at the central region of small amplitude oscillons
or boson stars is proportional to $\varepsilon^4$. On the other hand,
the cosmological constant can be interpreted as a fluid with energy
density,
\begin{equation}
\mu_\Lambda=\frac{\Lambda}{8\pi}   \label{mulambda}
\end{equation}
and pressure $p_\Lambda=-\mu_\Lambda$.  Using \eqref{caphlambda} and
\eqref{caphlowh} $\mu_\Lambda$ can be written as
\begin{equation}
\mu_\Lambda=\frac{D(D-1)}{16\pi}\varepsilon^4m^2h^2 \,. \label{mulambda2}
\end{equation}
The smallness of $h$ guaranties that $\mu_\Lambda$ remains smaller
than energy density $\mu$ of the scalar field even in the
small-amplitude limit.

We expand $\phi$ and the metric functions in powers of $\varepsilon$
as
\begin{align}
\phi&=\sum_{k=1}^\infty\epsilon^{2k}\phi_{2k} \,,\label{eq:phiexp}\\
A&=\frac{\Bigl(1-\frac{\varepsilon^2h^2\rho^2}{4}\Bigr)^2}
{\Bigl(1+\frac{\varepsilon^2h^2\rho^2}{4}\Bigr)^2}
+\sum_{k=1}^\infty\epsilon^{2k}A_{2k} \,, \label{eq:aexp}\\
B&=\frac{1}{\Bigl(1+\frac{\varepsilon^2h^2\rho^2}{4}\Bigr)^2}
+\sum_{k=1}^\infty\epsilon^{2k}B_{2k} \,. \label{eq:bexp}
\end{align}
The first terms in $A$ and $B$ represent the de Sitter background,
according to \eqref{desittermet}.  Since we intend to use
asymptotically de Sitter coordinates, we look for functions
$\phi_{2k}$, $A_{2k}$ and $B_{2k}$ that tend to zero when $\rho$ is
large.  One could initially include odd powers of $\varepsilon$ into
the expansions \eqref{eq:phiexp}-\eqref{eq:bexp}.  However, for
oscillatons it can be shown by the method presented below, that the
coefficients of those terms necessarily vanish when we are looking for
configurations that remain bounded in time. It can also be verified
directly, that the small-amplitude expansion of boson stars do not
include odd powers of $\varepsilon$. This is in contrast to flat
background oscillons, where the leading order behavior of the
amplitude is proportional to $\varepsilon$.

The frequency of boson stars and oscillatons also depends on their
amplitude. The smaller the amplitude is, the closer the frequency
becomes to the threshold $m$. Hence we introduce a rescaled time
coordinate $\tau$ by
\begin{equation}
\tau=\omega t \,,
\end{equation}
and expand the square of the $\varepsilon$ dependent factor $\omega$
as
\begin{equation}
\omega^2=m^2\left(1+\sum_{k=1}^\infty\varepsilon^{2k}
\omega_{2k} \right) .
\end{equation}
It is possible to allow odd powers of $\varepsilon$ into the expansion
of $\omega^2$, but the coefficients of those terms turn out to be zero
when solving the equations arising from the small-amplitude expansion.
There is a considerable freedom in choosing different parametrizations
of the small-amplitude states, changing the actual form of the
function $\omega$. The physical parameter is not $\varepsilon$ but the
frequency of the periodic states that will be given by
$\omega$. Similarly to the asymptotically flat case in
\cite{oscillaton} and to the dilaton model in \cite{dilaton}, it turns
out that for spatial dimensions $2<D<6$ the parametrization of the
small-amplitude states can be fixed by setting
$\omega=m\sqrt{1-\varepsilon^2}$.

The field equations we intend to solve using the $\tau$ and $\rho$
coordinates are the Einstein equations \eqref{eq:einst}, substituting
the Einstein tensor components from \eqref{eq:gtt}-\eqref{eq:gthth},
together with the wave equation \eqref{eq:wave1} or
\eqref{eq:wave2}. We note that these equations are not
independent. The $(\tau,\rho)$ component of the Einstein equations is
a constraint, and the wave equation is a consequence of the contracted
Bianchi identity.

\subsection{Boson stars} \label{secbsexp}

In case of boson stars the metric is static, hence $A_k$ and $B_k$ are
time independent. The complex scalar field $\phi$ oscillates with
frequency $\omega$,
\begin{equation}
\phi=\psi e^{i\omega t} \,,
\end{equation}
where $\psi$ is real and depends only on the radial coordinate $\rho$.
The small-amplitude expansion coefficients of $\phi$ satisfy
\begin{equation}
\phi_k=\psi_k e^{i\omega t}=\psi_k e^{i\tau} \,, \label{phipsi}
\end{equation}
where $\psi_k$ are real functions of $\rho$.  Substituting the
expansion \eqref{eq:phiexp}-\eqref{eq:bexp} into the field equations,
if $D\neq 2$, then to leading $\varepsilon^4$ order the $(\rho,\rho)$
component gives
\begin{equation}
B_2=\frac{A_2}{2-D} \,.
\end{equation}
Since for $D=2$ there is no solution representing a localized object
we assume $D>2$.  Then the $(\tau,\tau)$ component and the wave
equation yield to leading order a coupled system of differential
equations for $A_2$ and $\psi_2$,
\begin{align}
\frac{d^2A_2}{d\rho^2}+\frac{D-1}{\rho}\,\frac{dA_2}{d\rho}&=
2\frac{D-2}{D-1}\,\psi_2^2 \,, \label{eq:snbs1}\\
\frac{d^2\psi_2}{d\rho^2}+\frac{D-1}{\rho}\,\frac{d\psi_2}{d\rho}&=
\psi_2(A_2-\omega_2-h^2\rho^2) \,. \label{eq:snbs2}
\end{align}
Introducing the functions $s$ and $S$ by
\begin{equation}
s=\omega_2-A_2 \ , \quad S=\psi_2\sqrt{2\frac{D-2}{D-1}} \ ,
\label{eq:sandsbs}
\end{equation}
equations \eqref{eq:snbs1} and \eqref{eq:snbs2} can be written into
the form
\begin{align}
\frac{d^2S}{d\rho^2}
+\frac{D-1}{\rho}\,\frac{d S}{d\rho}
+(s+h^2\rho^2) S&=0
\,, \label{Seq0}\\
\frac{d^2s}{d\rho^2}
+\frac{D-1}{\rho}\,\frac{d s}{d\rho}
+S^2&=0
\,. \label{seq0}
\end{align}
Since the same equations will be obtained for small-amplitude
oscillatons in the next subsection, we postpone the discussion of this
system to Subsection \ref{sec:sn}.

Proceeding to higher orders, the $\varepsilon^6$ components give a
system of lengthy differential equations linear in $A_4$, $B_4$,
$\psi_4$ and their derivatives. These equations have nonlinear source
terms containing $A_2$, $B_2$ and $\psi_2$. They will also involve the
first coefficient $u_1$ from the series expansion of the potential
$U$. To leading order, the structure of small-amplitude boson stars
only depends on the mass parameter $m$.

\subsection{Oscillatons} \label{secoscexp}

In this case the scalar field is real, and the metric components are
also oscillating. We are looking for solutions that remain bounded as
time passes. It turns out, that these configurations are necessarily
periodically oscillating in time.

From the $\varepsilon^2$ components of the field equations follows
that $B_2$ only depends on $\rho$ and that
\begin{equation}
\phi_2=p_2\cos(\tau+\delta)  \,, \label{phi2p2}
\end{equation}
where two new functions, $p_2$ and $\delta$ are introduced, depending
only on $\rho$. From the $\varepsilon^4$ part of the $(\tau,\rho)$
component of the Einstein equations it follows that $B_4$ can remain
bounded as time passes only if $\delta$ is a constant. Then by a shift
in the time coordinate we set
\begin{equation}
\delta=0 \,.
\end{equation}
This shows that the scalar field oscillates simultaneously, with the
same phase at all radii.

From the $\varepsilon^4$ component of the difference of the
$(\rho,\rho)$ and $(\theta_1,\theta_1)$ Einstein equations follows
that $A_2$ also depends only on the radial coordinate $\rho$, showing
that the metric is static to order $\varepsilon^2$. This is the main
advantage of using isotropic coordinates over Schwarzschild
coordinates. But even in this system the coefficients $A_4$ and $B_4$
will already contain $\cos(2\tau)$ terms. From the $\varepsilon^4$
component of the $(\tau,\rho)$ Einstein equations follows that
\begin{equation}
B_4=b_4-\frac{p_2^2}{4(D-1)}\cos(2\tau) \,,
\end{equation}
where $b_4$ is a function of $\rho$. The $(\rho,\rho)$ component shows
that if $D\not=2$ then
\begin{equation}
B_2=\frac{A_2}{2-D} \,.
\end{equation}
If $D=2$ then $A_2=0$, and there are no nontrivial localized regular
solutions for $B_2$ and $p_2$, so we assume $D>2$ from now. The
$(\tau,\tau)$ component yields
\begin{equation}
\frac{d^2A_2}{d\rho^2}+\frac{D-1}{\rho}\,\frac{dA_2}{d\rho}=
\frac{D-2}{D-1}\,p_2^2 \,. \label{eq:a2osc}
\end{equation}
The wave equation provides an equation that determines the time
dependence of $\phi_4$. In order to keep $\phi_4$ bounded, the
resonance terms proportional to $\cos\tau$ must vanish, yielding
\begin{equation}
\frac{d^2p_2}{d\rho^2}+\frac{D-1}{\rho}\,\frac{dp_2}{d\rho}=
p_2(A_2-\omega_2-h^2\rho^2) \,. \label{eq:p2osc}
\end{equation}

Equations \eqref{eq:a2osc} and \eqref{eq:p2osc} do not depend on the
coefficients $\bar u_k$ of the potential $\bar U(\phi)$. This means
that the leading order small-amplitude behavior of oscillatons is
always the same as for the Klein-Gordon case.

Introducing the functions $s$ and $S$ by
\begin{equation}
s=\omega_2-A_2 \ , \quad
S=p_2\sqrt{\frac{D-2}{D-1}} \ , \label{ssa2p2osc}
\end{equation}
equations \eqref{eq:a2osc} and \eqref{eq:p2osc} can be written into
the form \eqref{Seq0} and \eqref{seq0} already obtained at the
small-amplitude expansion of boson stars.

\subsection{Schr\"odinger-Newton equations} \label{sec:sn}

The equations describing both small-amplitude boson stars and
oscillatons on an expanding background has been written into a form,
which for $h=0$ reduces to the time-independent Schr\"odinger-Newton
(SN) equations \cite{Diosi,Penrose,Moroz,Tod}
\begin{align}
\frac{d^2S}{d\rho^2}
+\frac{D-1}{\rho}\,\frac{d S}{d\rho}
+(s+h^2\rho^2) S&=0
\,, \label{Seq}\\
\frac{d^2s}{d\rho^2}
+\frac{D-1}{\rho}\,\frac{d s}{d\rho}
+S^2&=0
\,. \label{seq}
\end{align}

If $S(\rho)$ and $s(\rho)$ are solutions of \eqref{Seq} and
\eqref{seq}, then the transformed functions
\begin{equation}
\tilde S(\rho)=\lambda^2S(\lambda\rho) \ ,\quad
\tilde s(\rho)=\lambda^2s(\lambda\rho)  \label{eq:sntr}
\end{equation}
solve the equations
\begin{align}
\frac{d^2\tilde S}{d\rho^2}
+\frac{D-1}{\rho}\,\frac{d\tilde S}{d\rho}
+(\tilde s+\tilde h^2\rho^2)\tilde S&=0
\,, \label{Seqt}\\
\frac{d^2\tilde s}{d\rho^2}
+\frac{D-1}{\rho}\,\frac{d\tilde s}{d\rho}
+\tilde S^2&=0
\,, \label{seqt}
\end{align}
which are obtained from the SN equations by replacing the constant $h$
by $\tilde h=\lambda^2h$.

If the scalar field, and consequently $S$, tends to zero for
$\rho\to\infty$ then \eqref{seq} gives the following approximation for
$s$,
\begin{equation}
s=s_0+s_1\rho^{2-D} \,.  \label{sasympt}
\end{equation}
Since we are looking for solutions for which $A_2=\omega_2-s$ also
approaches zero for large $\rho$, necessarily $\omega_2=s_0$.
In order to make the solution of the SN equations
\eqref{Seq}-\eqref{seq} unique we demand
\begin{equation}
\omega_2=s_0=-1.
\end{equation}
Setting $\omega_k=0$ for $k\geq3$ this also fixes the connection
between the frequency $\omega$ and the small-amplitude parameter
$\varepsilon$, giving
\begin{equation}
\omega=m\sqrt{1-\varepsilon^2} \,. \label{omegaepsilon}
\end{equation}

Substituting into \eqref{Seq} we get
\begin{equation}
\frac{d^2S}{d\rho^2}
+\frac{D-1}{\rho}\,\frac{d S}{d\rho}
+(h^2\rho^2-1+s_1\rho^{2-D}) S=0
\,, \label{swkb1}
\end{equation}
determining the large $\rho$ behavior of $S$.

\subsection{Energy density} \label{sec:enden}

The energy density of the scalar field is $\mu=T_{ab}u^au^b$, where
the unit timelike vector $u^a$ has the components
$(1/\sqrt{A},0,...,0)$.  In terms of the rescaled scalar field $\phi$,
the energy density of oscillatons can be written as
\begin{equation}
\mu=\frac{1}{16\pi}\left[\frac{1}{A}\left(\frac{d\phi}{dt}\right)^2
+\frac{1}{B}\left(\frac{d\phi}{dr}\right)^2+2\bar U(\phi)\right] ,
\label{eq:massenosc}
\end{equation}
and for boson stars
\begin{equation}
\mu=\frac{1}{16\pi}\left[
\frac{1}{A}\frac{d\phi^*}{dt}\frac{d\phi}{dt}
+\frac{1}{B}\frac{d\phi^*}{dr}\frac{d\phi}{dr}
+\bar U(\phi^*\phi)\right] .
\label{eq:massenbs}
\end{equation}
In case of boson stars, using \eqref{eq:phiexp}, \eqref{phipsi} and
\eqref{eq:sandsbs}, to leading order in $\varepsilon$ we get
\begin{equation}
\mu=\frac{1}{16\pi}\varepsilon^4\frac{D-1}{D-2}m^2S^2 \,.
\label{massenlead}
\end{equation}
For oscillatons we obtain the same expression for $\mu$ using
\eqref{eq:phiexp}, \eqref{phi2p2} and \eqref{ssa2p2osc}. Measuring the
scalar field mass $m$ in $eV/c^2$ units, the energy density at the
symmetry center for $D=3$ can be written as
\begin{equation}
\mu_c=\varepsilon^4\left(\frac{mc^2}{eV}\right)^2
1.436\times 10^{39}\frac{kg}{\mathrm{m}^3}\,. \label{massendens}
\end{equation}
(We use Roman $\mathrm{m}$ for meters in order to distinguish from the
scalar field mass $m$.)  Here we have used the numerically obtained
central value of $S$, which is $S_c=1.0215$ for $D=3$.

The effective energy density $\mu_\Lambda$ corresponding to the
present value of the cosmological constant is given in ordinary units
by \eqref{cosmdens} in Appendix \ref{sec:cosmconst}.  Comparing to
\eqref{massendens}, one can see that the energy density of the scalar
field is much higher than $\mu_\Lambda$ unless $\varepsilon$ or $m$ is
extremely small.  Using \eqref{mulambda2}, the ratio of $\mu_\Lambda$
to the central density $\mu_c$ can be written as
\begin{equation}
\frac{\mu_\Lambda}{\mu_c}=\frac{D(D-2)}{S_c^2}h^2 \,, \label{mumulambda}
\end{equation}
providing an important physical interpretation for the rescaled Hubble
constant $h$. In Table \ref{sctable} we give the value of $S_c$ and
the coefficient of $h^2$ in $\mu_\Lambda/\mu_c$ for the relevant
dimensions.
\begin{table}[htbp]
\begin{tabular}{|c|c|c|c|}
\hline
  & $D=3$  & $D=4$  & $D=5$ \\
  \hline
  $S_c$
  & $1.0215$ & $3.5421$ & $14.020$ \\
  $D(D-2)/S^2_c$
  & $2.8751$ & $0.63761$ & $0.076313$ \\
  \hline
\end{tabular}
\caption{The numerically calculated values of the function $S$ at the
center and the coefficient of $h^2$ in $\mu_\Lambda/\mu_c$
for $D=3$, $4$ and $5$ spatial dimensions.  \label{sctable}}
\end{table}
We will see in the following sections that $h$ turns out to be the
essential parameter determining the energy loss rate of boson stars
and oscillatons.

\subsection{Outer core region}

If $2<D<6$, and the cosmological constant is zero, i.e.~$h=0$, then
assuming that $s_0=-1$ the SN equations have a unique localized
nodeless solution. Solutions with nodes have higher energy and are
unstable.  For large $\rho$ the function $S$ satisfies the equation
\begin{equation}
\frac{d^2S}{d\rho^2}
+\frac{D-1}{\rho}\,\frac{d S}{d\rho}
+(-1+s_1\rho^{2-D}) S=0
\,, \label{outcores}
\end{equation}
which has exponentially decaying solutions
\begin{align}
S&=S_t\frac{e^{-\rho}}{\rho^{1-s_1/2}}\left[1
+\mathcal{O}\left(\frac{1}{\rho}\right)\right] , &
D&=3 \,, \label{smid3d} \\
S&=S_t\frac{e^{-\rho}}{\rho^{(D-1)/2}}\left[1
+\mathcal{O}\left(\frac{1}{\rho}\right)\right] , &
D&>3 \,. \label{smid45d}
\end{align}
The numerically determined values of the constants $s_1$ and $S_t$ for
the case $h=0$ are given in Table \ref{c1table}.
\begin{table}[htbp]
\begin{tabular}{|c|c|c|c|}
\hline
  & $D=3$  & $D=4$  & $D=5$ \\
  \hline
  $s_1$ & $3.50533$ & $7.69489$ & $10.4038$ \\
  $S_t$ & $3.49513$ & $88.2419$ & $23.3875$ \\
  \hline
\end{tabular}
\caption{The numerical values of the constants $s_1$ and $S_t$
  for $h=0$ in $3$, $4$ and $5$ spatial dimensions. \label{c1table}}
\end{table}

Even if $h$ is nonzero, we assume that it is small enough such that
there is a region of $\rho$ which is well outside the core region, but
where the influence of the cosmological constant is still
negligible. In this region $S$ satisfies \eqref{outcores}, and
consequently \eqref{smid3d} and \eqref{smid45d} are good
approximations together with the values of the constants belonging to
the $h=0$ case given in Table \ref{c1table}.

\subsection{Oscillating tail region}

If $h$ is nonzero, at very large distances $\rho\gg 1/h$, and the
behavior of $S$ is determined by the equation
\begin{equation}
\frac{d^2S}{d\rho^2}
+\frac{D-1}{\rho}\,\frac{d S}{d\rho}+h^2\rho^2 S=0 \,.
\end{equation}
The function $S$ has an oscillating standing wave tail in this
domain,
\begin{equation}
S=\frac{\alpha}{\rho^{\frac{D}{2}}}
\cos\left(\frac{h\rho^2}{2}+\beta\right) , \label{stail}
\end{equation}
where $\alpha$ and $\beta$ are constants, describing the amplitude and
phase.  We can interpret this standing wave as the superposition of an
outgoing wave carrying energy out from the central object and an
artificial ingoing wave component that is added to keep the solution
exactly periodic. We will relate the amplitude of the tail to the
energy loss rate of dynamical boson stars or oscillatons.

\begin{figure}[!ht]
    \begin{center}
    \includegraphics[width=8.6cm]{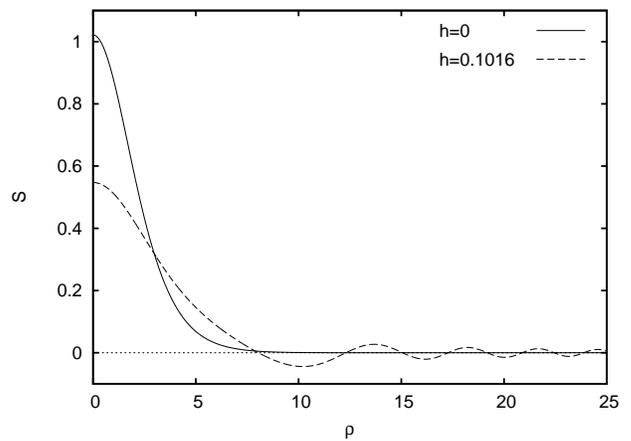}
    \end{center}
    \caption{The radial dependence of the function $S$ for a large
      value of $h$ is compared to the $h=0$ flat background case. The
      function $S$ for $h=0$ decays exponentially, while for $h>0$
      curve starts to oscillate. For $h=0.1016$ the first change in
      signature is at $\rho_t=8.04$.
      \label{splot}}
\end{figure}

When solving the SN equations \eqref{Seq} and \eqref{seq} by a
numerical method, the amplitude $\alpha$ depends on the central values
of $s$ and $S$ given at $\rho=0$.  We intend to determine the minimal
tail amplitude, $\alpha=\alpha_{\mathrm{min}}$, assuming that $s$
tends to $s_0=-1$ for $\rho\to\infty$. We are interested in solutions
which has no nodes in the central region where the influence of $h$ is
negligible. If $h>0$ then $\alpha_{\mathrm{min}}$ is nonzero.
In order to show the influence of $h$ on the form of the function $S$,
on Fig.~\ref{splot} we plot the radial dependence of $S$ for
$h=0.1016$ and for $h=0$.

Assuming that $h$ is small, we calculate the amplitude $\alpha$ by WKB
analysis.  Our aim is to continue the function $S$ given by
\eqref{smid3d} and \eqref{smid45d} into the very large distances where
the cosmological term $h^2\rho^2$ dominates.  We will express the
amplitude $\alpha$ in terms of the constant $S_t$. To the function
which is exponentially decaying when going outwards at the outer core
region belongs an oscillating tail with a certain amplitude and
phase. Adding the mode which exponentially grows when going outwards
does not change the function noticeably in the outer core region but
adds an oscillating tail with a phase shift $\pi/2$. The resulting
tail amplitude is minimal if this second mode is not present. Hence
the WKB method will give the minimum amplitude
$\alpha=\alpha_{\mathrm{min}}$.  In Sec. \ref{sntailsec} we will
compare the WKB amplitude to results obtained by direct numerical
integration of the SN equations. Besides indicating the correctness of
the WKB results, this will also give information on how large $h$ can
be chosen such that the WKB result can be still considered a valid
approximation.

\section{WKB formalism}

\subsection{Schr\"odinger form}

Introducing a new radial coordinate
\begin{equation}
y=h\rho
\end{equation}
we transform \eqref{swkb1} into a form corresponding to a
Schr\"odinger equation with $h$ playing the role of the Planck's
constant $\hbar$,
\begin{equation}
h^2\left(\frac{d^2S}{dy^2}
+\frac{D-1}{y}\,\frac{d S}{dy}\right)
+\left(y^2-1+\frac{h^{D-2}}{y^{D-2}}s_1\right) S=0 \,.
\end{equation}
Introducing
\begin{equation}
S=y^{(1-D)/2}Z \,,  \label{szyeq}
\end{equation}
we can transform this into a one-dimensional form,
\begin{equation}
h^2\frac{d^2Z}{dy^2}+p^2Z=0 \,, \label{onedsch}
\end{equation}
where
\begin{equation}
p^2=y^2-1+\frac{h^{D-2}}{y^{D-2}}s_1-\frac{h^2}{4y^2}(D-1)(D-3) \,.
\end{equation}
The outer domain of the core region is represented now by $0<y\ll 1$,
where
\begin{align}
Z&=S_th^{1-s_1/2}y^{s_1/2}e^{-y/h}\left[1
+\mathcal{O}\left(\frac{h}{y}\right)\right] , &
D&=3 \,, \label{zsy3d} \\
Z&=S_th^{(D-1)/2}e^{-y/h}\left[1
+\mathcal{O}\left(\frac{h}{y}\right)\right] , &
D&>3 \,. \label{zsy45d}
\end{align}

\subsection{WKB approximation}

Introducing a new function $f$ by
\begin{equation}
Z=e^{if/h}  \label{zeq}
\end{equation}
and substituting into \eqref{onedsch} we get
\begin{equation}
ih\frac{d^2f}{dy^2}-\left(\frac{df}{dy}\right)^2+p^2=0 \,.
\label{onedsch2}
\end{equation}
We write f as a power series in $h$,
\begin{equation}
f=\sum_{k=0}^\infty f_kh^k \,.
\end{equation}
Differently from the standard WKB approach, now $p^2$ also has to be
expanded in $h$,
\begin{equation}
p^2=\sum_{k=0}^\infty p_k^2h^k \,,
\end{equation}
where the nonvanishing coefficients are
\begin{align}
p_0^2&=y^2-1 \,, \label{p02}\\
p_2^2&=-\frac{(D-1)(D-3)}{4y^2} \,, \\
p_{D-2}^2&=\frac{s_1}{y^{D-2}}  \,.
\end{align}
Substituting into \eqref{onedsch2}, to leading order we get
\begin{equation}
\left(\frac{df_0}{dy}\right)^2=p_0^2 \,, \qquad
\frac{df_0}{dy}=\pm p_0 \,.
\end{equation}
The order $h$ components yield
\begin{equation}
i\frac{d^2f_0}{dy^2}+p_1^2=2\frac{df_0}{dy}\,\frac{df_1}{dy}
\label{hcomp}\,.
\end{equation}

\subsection{WKB for $D>3$}

If $D>3$ then $p_1^2=0$,
\begin{equation}
f_1=\frac{i}{2}\ln\left(\frac{df_0}{dy}\right)
=\frac{i}{2}\ln\left(\pm p_0\right) \,,
\end{equation}
and we get the standard WKB
result,
\begin{align}
Z&=\frac{A_\pm}{\sqrt{|p_0|}}
\exp\left(\pm\frac{i}{h}\int_1^y p_0dy\right) , & 0&<y<1
\,,\label{z45da} \\
Z&=\frac{B_\pm}{\sqrt{p_0}}
\exp\left(\pm\frac{i}{h}\int_1^y p_0dy\right) , & y&>1
\,.\label{z45db}
\end{align}
The general solution for $0<y<1$ is a sum of two terms with
proportionality constants $A_+$ and $A_-$, and similarly for the $y>1$
case with constants $B_+$ and $B_-$. For the $0<y<1$ case
$|p_0|=\sqrt{1-y^2}$. Substituting \eqref{p02}, it is possible to
perform the integral in the exponentials. For $0<y<1$,
\begin{align}
\frac{i}{h}\int_1^y p_0 dy&=
-\frac{1}{h}\int_1^y\sqrt{1-y^2}\,dy \notag\\
&=-\frac{1}{2h}\left(y\sqrt{1-y^2}
+\arcsin y-\frac{\pi}{2}\right) ,  \label{p0int1}
\end{align}
while for $y>1$,
\begin{align}
\frac{i}{h}\int_1^y p_0 dy&=
\frac{i}{h}\int_1^y\sqrt{y^2-1}\,dy \notag\\
&=\frac{i}{2h}\left[y\sqrt{y^2-1}
-\ln\left(y+\sqrt{y^2-1}\right)\right] . \label{p0int2}
\end{align}
In the region $0<y\ll 1$, substituting the expansion of
\eqref{p0int1}, to leading order \eqref{z45da} takes the form
\begin{equation}
Z=A_\pm\exp\left[\mp\frac{1}{h}\left(y-\frac{\pi}{4}\right)\right] .
\end{equation}
Comparing with \eqref{zsy45d}, follows that for $D>3$
\begin{equation}
A_+=S_th^{(D-1)/2}\exp\left(-\frac{\pi}{4h}\right) \,, \qquad
A_-=0 \,.  \label{apd45}
\end{equation}

\subsection{WKB for $D=3$}

Because of the term proportional to $h$ in $p^2$, the $D=3$ case
has to be treated separately. Then \eqref{hcomp}
takes the form
\begin{equation}
\frac{i}{2}\frac{d}{dy}
\left[\ln\left(\pm p_0\right)\right]
\pm\frac{s_1}{2yp_0}=\frac{df_1}{dy} \,.
\end{equation}
Using \eqref{p02},
\begin{equation}
\frac{i}{2}\frac{d}{dy}
\left[\ln\left(\pm\sqrt{y^2-1}\right)\right]
\pm\frac{s_1}{2iy\sqrt{1-y^2}}=\frac{df_1}{dy} \,.
\end{equation}
Integrating, for $0<y<1$,
\begin{align}
f_1&=\frac{i}{2}\ln\left(\sqrt{y^2-1}\right) \\
&\quad\pm\frac{s_1}{2i}\left[\ln y
-\ln\left(1+\sqrt{1-y^2}\right)\right]
+c_1 \,, \notag
\end{align}
and for $y>1$,
\begin{align}
f_1&=\frac{i}{2}\ln\left(\sqrt{y^2-1}\right) \\
&\quad\mp\frac{s_1}{2}\arctan\left(\frac{1}{\sqrt{y^2-1}}\right)
+c_2 \,. \notag
\end{align}
Substituting into \eqref{zeq}, for $0<y<1$,
\begin{equation}
Z=\frac{A_\pm}{\sqrt{|p_0|}}
\left(\frac{y}{1+\sqrt{1-y^2}}\right)^{\pm s_1/2}
\exp\left(\pm\frac{i}{h}\int_1^y p_0 dy\right) , \label{zs1d3}
\end{equation}
and for $y>1$,
\begin{align}
Z&=\frac{B_\pm}{\sqrt{p_0}}
\exp\left\{\mp\frac{is_1}{2}\left[\arctan
\left(\frac{1}{\sqrt{y^2-1}}\right)-\frac{\pi}{2}
\right]\right\} \notag\\
&\qquad\times\exp\left(\pm\frac{i}{h}\int_1^y p_0 dy
\right) . \label{zs1d45}
\end{align}
For $0<y\ll 1$ the expansion of \eqref{zs1d3} gives
\begin{equation}
Z=A_\pm \left(\frac{y}{2}\right)^{\pm s_1/2}
\exp\left[\mp\frac{1}{h}\left(y-\frac{\pi}{4}\right)\right] .
\end{equation}
Comparing with \eqref{zsy3d}, follows that for $D=3$
\begin{equation}
A_+=2^{s_1/2}S_th^{1-s_1/2}\exp\left(-\frac{\pi}{4h}\right) \,, \qquad
A_-=0 \,. \label{apd3}
\end{equation}

\subsection{Connection formulae}

The next step is to relate the amplitudes $A_\pm$ in the $y<1$ region
to the amplitudes $B_\pm$ for $y>1$. The expressions \eqref{z45da} and
\eqref{z45db} for $D>3$ are just the standard WKB solutions, so one
can apply the appropriate connection formulae to express $B_\pm$ in
terms of $A_\pm$. In our case we need the one where the field is
exponentially increases when going away from the turning point (see
for example (34.18) of \cite{schiff})
\begin{align}
&\frac{1}{\sqrt{p_0}}
\cos\left(\frac{1}{h}\int_1^y p_0dy+\frac{\pi}{4}\right) \notag\\
&\qquad\longrightarrow
\frac{1}{\sqrt{|p_0|}}
\exp\left(\frac{1}{h}\int_y^1 |p_0|dy\right)
\,,
\end{align}
from $y>1$ to $y<1$. This is equivalent to
\begin{equation}
A_-=0 \ , \quad
B_\pm=\frac{A_+}{2}
\exp\left(\pm\frac{i\pi}{4}\right) . \label{bpmap}
\end{equation}
The expressions \eqref{zs1d3} and \eqref{zs1d45} for $D=3$ differ from
the standard WKB solutions by factors involving $s_1$. However, since
both of these factors take the value $1$ at the $y=1$ turning point,
the formulae \eqref{bpmap} relating the amplitudes at the two sides
hold in this case too. A detailed derivation of \eqref{bpmap} is given
in Appendix \ref{app:con}.

\subsection{Schr\"odinger-Newton tail amplitude}\label{sntailsec}

For $y\gg1$ the expression \eqref{p0int2} can be approximated by
\begin{equation}
\frac{i}{h}\int_1^y p_0 dy=\frac{iy^2}{2h} \,.
\end{equation}
In the $D>3$ case to leading order \eqref{z45db} yields
\begin{equation}
Z=\frac{B_\pm}{\sqrt{y}}\exp\left(\pm\frac{iy^2}{2h}\right) .
\end{equation}
Substituting $B_\pm$ from \eqref{bpmap} gives
\begin{equation}
Z=\frac{A_+}{\sqrt{y}}\cos\left(\frac{y^2}{2h}+\frac{\pi}{4}\right)
\ , \qquad D>3 \ .
\end{equation}
For $D=3$, to leading order \eqref{zs1d45} yields
\begin{equation}
Z=\frac{B_\pm}{\sqrt{y}}\exp
\left[\pm i\left(\frac{y^2}{2h}+\frac{\pi s_1}{4}\right)\right] ,
\end{equation}
and substituting $B_\pm$ from \eqref{bpmap} gives
\begin{equation}
Z=\frac{A_+}{\sqrt{y}}\cos\left[\frac{y^2}{2h}
+\frac{\pi}{4}(s_1+1)\right]
\ , \qquad D=3 \  .
\end{equation}

Using \eqref{szyeq} we can express the original $S$ variable in the SN
equations using the coordinate $\rho=y/h$, obtaining the same
asymptotic formula as in \eqref{stail},
\begin{equation}
S=\frac{\alpha}{\rho^{\frac{D}{2}}}
\cos\left(\frac{h\rho^2}{2}+\beta\right) , \label{stail2}
\end{equation}
where the amplitude is
\begin{equation}
\alpha=\frac{A_+}{h^{D/2}}
\end{equation}
for any $D$, and the phase is
\begin{equation}
\beta=\left\{
\begin{array}{ll}
\dfrac{\pi}{4}(s_1+1) &\text{if } D=3 \ , \\[3mm]
\dfrac{\pi}{4} &\text{if } D>3 \ .
\end{array}
\right.
\end{equation}

Substituting the value of $A_+$ from \eqref{apd45} and \eqref{apd3}
gives our final WKB result for the minimal amplitude
\begin{equation}
\alpha_{\mathrm{min}}=\left\{
\begin{array}{ll}
\displaystyle{\frac{S_t}{\sqrt{h}}
\left(\frac{2}{h}\right)^{\frac{s_1}{2}}
\exp\left(-\frac{\pi}{4h}\right)}
&\text{ if  } D=3 \ , \\[4mm]
\displaystyle{\frac{S_t}{\sqrt{h}}\exp\left(-\frac{\pi}{4h}\right)}
&\text{ if  } D>3 \ .
\end{array}
\right.  \label{alphamin}
\end{equation}

\begin{figure}[!ht]
    \begin{center}
    \includegraphics[width=8.6cm]{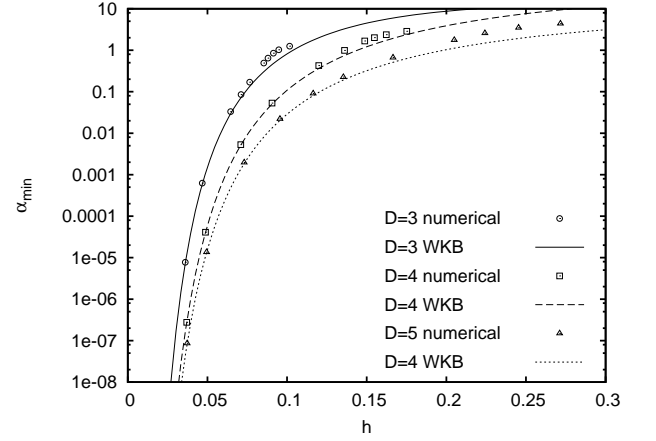}
    \end{center}
    \caption{The curves represent the WKB tail amplitude, while the
      points the numerically calculated values, as a function of $h$,
      for $D=3,4,5$ spatial dimensions.
      \label{figampl}}
\end{figure}

\begin{figure}[!ht]
    \begin{center}
    \includegraphics[width=8.6cm]{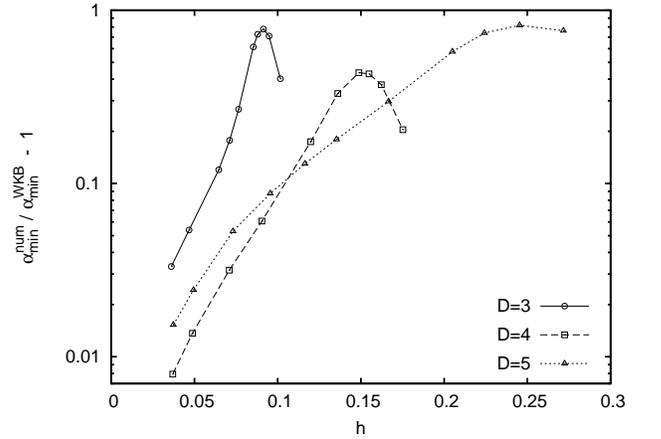}
    \end{center}
    \caption{Relative difference of the numerically calculated minimal
      tail amplitude from the analytically calculated WKB values.
      \label{figdiff}}
\end{figure}

It is instructive to compare the above analytical result to the
minimal amplitude obtained by the numerical solution of the SN
equations \eqref{Seq}-\eqref{seq}. For a chosen $h$ one has to
minimize the oscillating tail for those solutions for which the field
$s$ tends to $s_0=-1$ asymptotically, while $S$ has no nodes in the
core region. The solutions depend very strongly on the central values
$s_c$ and $S_c$. The results for various $h$ for $D=3$ spatial
dimensions are listed in Table \ref{d3amptab}.
\begin{table}[htbp]
\begin{tabular}{|c|c|c|c|c|c|}
\hline
$h$  & $s_c/s_c^{h=0}$  & $S_c/S_c^{h=0}$ & $\rho_t$
& $\alpha_{\mathrm{min}}^{\mathrm{num}}$
& $\frac{\alpha_{\mathrm{min}}^{\mathrm{num}}}{
\alpha_{\mathrm{min}}^{\mathrm{WKB}}}-1$ \\
  \hline
$0.036111$ & $0.96295$ & $0.97043$ & $28.65$ & $7.7377\times10^{-6}$ & $0.033$\\
$0.046769$ & $0.93539$ & $0.94830$ & $22.04$ & $6.2645\times10^{-4}$ & $0.054$\\
$0.064593$ & $0.86123$ & $0.88802$ & $15.73$ & $0.033101$ & $0.120$\\
$0.071112$ & $0.81742$ & $0.85171$ & $14.13$ & $0.085425$ & $0.177$\\
$0.076502$ & $0.76580$ & $0.80827$ & $12.90$ & $0.16998$ & $0.268$\\
$0.085350$ & $0.63180$ & $0.69420$ & $10.62$ & $0.49007$ & $0.614$\\
$0.088000$ & $0.58803$ & $0.65720$ & $9.94$  & $0.64553$ & $0.727$\\
$0.091400$ & $0.53709$ & $0.61443$ & $9.24$  & $0.85085$ & $0.778$\\
$0.094800$ & $0.49518$ & $0.57976$ & $8.72$  & $1.02507$ & $0.709$\\
$0.101600$ & $0.43960$ & $0.53576$ & $8.04$  & $1.25256$ & $0.402$\\
  \hline
\end{tabular}
\caption{For several choices of $h$ the numerically calculated minimal
  tail amplitude $\alpha_{\mathrm{min}}^{\mathrm{num}}$ and its relative
  difference from the WKB value \eqref{alphamin} is listed
  assuming $D=3$.
  We also give the ratio of central values of the functions $s$
  and $S$ to the flat background values
  $s_c^{h=0}=0.93832284$ and $S_c^{h=0}=1.02149304$. In order to
  show where the tail region starts, we also list the radius
  $\rho_t$, where $S$ first crosses the value zero.
  \label{d3amptab}}
\end{table}
The amplitudes $\alpha_{\mathrm{min}}$ for $D=3,4,5$ dimensions are
plotted on Fig.~\ref{figampl}, and the relative difference from the
analytically obtained WKB value is displayed on Fig.~\ref{figdiff}.
From this figure we read off that the WKB analysis gives less than
$10\%$ error, if $h<0.06$ in case of $D=3$, and if $h<0.1$ when $D=4$
or $5$.

\subsection{Scalar field tail amplitude}

For oscillatons, using \eqref{eq:phiexp}, \eqref{phi2p2},
\eqref{ssa2p2osc} and \eqref{stail2}, the real scalar in the tail
region behaves like
\begin{align}
\phi&=\varepsilon^2\phi_2=\varepsilon^2p_2\cos\tau
=\varepsilon^2 S\sqrt{\frac{D-1}{D-2}}\cos\tau \notag\\
&=\frac{\phi_A}{r^{\frac{D}{2}}}\cos(\omega t)
\cos\left(\frac{h\varepsilon^2m^2r^2}{2}+\beta\right) , \label{phitosc}
\end{align}
where the amplitude is
\begin{equation}
\phi_A=\frac{\varepsilon^{2-\frac{D}{2}}}{m^{\frac{D}{2}}}
\sqrt{\frac{D-1}{D-2}}\,
\alpha_{\mathrm{min}} \,. \label{phiaa}
\end{equation}
Applying the identity
\begin{equation}
\cos a\cos b=\frac{1}{2}\left[\cos(a-b)+\cos(a+b)\right] ,
\end{equation}
it is apparent that \eqref{phitosc} is a sum of an ingoing and
outgoing wave, both with amplitude $\phi_A/2$. One might be tempted to
add an ingoing wave with an opposite amplitude to obtain a purely
outgoing solution, and to conclude that the remaining outgoing
component has the amplitude $\phi_A/2$. However, an ingoing wave
contains a $\cos\left(h\varepsilon^2m^2r^2/2+\beta\right)$ component,
which would change the amplitude of the WKB mode which is increasing
when going away from the turning point. What one can do instead is
adding a standing wave proportional to
$\sin\left(h\varepsilon^2m^2r^2/2+\beta\right)$, corresponding to the
suppressed WKB mode. If the time dependence is also phase shifted,
then we obtain the minimal amplitude outgoing wave,
\begin{equation}
\phi=\frac{\phi_A}{r^{\frac{D}{2}}}
\cos\left(\frac{h\varepsilon^2m^2r^2}{2}+\beta-\omega t\right) .
\label{outwaosc}
\end{equation}

In case of boson stars, using \eqref{eq:phiexp}, \eqref{phipsi},
\eqref{eq:sandsbs} and \eqref{stail2}, the complex scalar in the tail
region behaves like
\begin{align}
\phi&=\varepsilon^2\phi_2=\varepsilon^2\psi_2e^{i\tau}
=\varepsilon^2 \frac{S}{\sqrt{2}}
\sqrt{\frac{D-1}{D-2}}\,e^{i\tau} \notag\\
&=\frac{\phi_A}{\sqrt{2}\,r^{\frac{D}{2}}}e^{i\omega t}
\cos\left(\frac{h\varepsilon^2m^2r^2}{2}+\beta\right) .
\end{align}
Now one can obtain the minimal amplitude outgoing wave by adding a
term
\begin{equation}
-i \frac{\phi_A}{\sqrt{2}\,r^{\frac{D}{2}}}e^{i\omega t}
\sin\left(\frac{h\varepsilon^2m^2r^2}{2}+\beta\right) ,
\end{equation}
obtaining
\begin{equation}
\phi=\frac{\phi_A}{\sqrt{2}\,r^{\frac{D}{2}}}
\exp\left[-i\left(\frac{h\varepsilon^2m^2r^2}{2}
+\beta-\omega t\right)\right] . \label{outwabs}
\end{equation}

\subsection{Mass loss rate}

Even if $\rho\gg 1/h$ in the oscillating tail region, according to
\eqref{eq:aexp} and \eqref{eq:bexp}, for small $\varepsilon$ the
metric functions $A$ and $B$ are still very close to $1$.  The mass
loss rate of the system can be calculated by applying the expression
\eqref{masstder} for the time derivative of the mass function from
Appendix \ref{appmass},
\begin{equation}
\frac{d\hat m}{dt}=\frac{2\pi^{\frac{D}{2}}r^{D-1}}
{\Gamma\left(\frac{D}{2}\right)}T_{tr} \,.
\end{equation}
Substituting the stress energy tensor from \eqref{eq:tab1} or
\eqref{eq:tab2},
\begin{align}
\frac{d\hat m}{dt}&=\frac{\pi^{\frac{D}{2}}r^{D-1}}
{\Gamma\left(\frac{D}{2}\right)}
\left(\Phi_{,t}^*\Phi_{,r}+\Phi_{,r}^*\Phi_{,t}\right) \notag\\
&=\frac{\pi^{\frac{D}{2}-1}r^{D-1}}
{8\Gamma\left(\frac{D}{2}\right)}
\left(\phi_{,t}^*\phi_{,r}+\phi_{,r}^*\phi_{,t}\right)
\,,  \label{eq:minkmt}
\end{align}
which expression is valid for both real and complex fields. We have to
evaluate this expression at large $r$ in the tail region. Substituting
the outgoing wave form \eqref{outwaosc} and averaging for one
oscillation period, or substituting \eqref{outwabs}, for both the
oscillaton and boson star case we obtain
\begin{equation}
\frac{d\hat m}{dt}=-
\frac{\pi^{\frac{D}{2}-1}h\varepsilon^2m^3\phi_A^2}
{8\Gamma\left(\frac{D}{2}\right)} \,.
\end{equation}
Here, since we are interested in the leading $\varepsilon$ order
result, we substituted $\omega=m$, and for large $r$ we neglected the
term arising when taking the derivative of the $r^{-D/2}$ factor.  For
large radius $\hat m$ agrees with the total mass $M$ of the oscillon
or boson star. Using \eqref{alphamin} and \eqref{phiaa} for the mass
loss rate and substituting $h=H/(m\varepsilon^2)$, for $D>3$ we obtain
\begin{equation}
\frac{dM}{dt}=-\frac{\pi^{\frac{D}{2}-1}(D-1)S_t^2\varepsilon^{6-D}}
{8m^{D-3}(D-2)\Gamma\left(\frac{D}{2}\right)}
\exp\left(-\frac{\pi m\varepsilon^2}{2H}\right) .
\end{equation}
If $D=3$ there is an extra term involving $s_1$,
\begin{equation}
\frac{dM}{dt}=-\frac{1}{2}S_t^2\varepsilon^3
\left(\frac{2m\varepsilon^2}{H}\right)^{s_1}
\exp\left(-\frac{\pi m\varepsilon^2}{2H}\right) .  \label{masslossd3}
\end{equation}

For small-amplitude configurations the parameter $\varepsilon$ can be
expressed by the mass $M$ using \eqref{meps}. Because of the
complexity of the resulting expression we only give the result for
$D=3$, when $M=\varepsilon s_1/(2m)$
\begin{equation}
\frac{dM}{dt}=-4S_t^2\frac{m^3M^3}{s_1^3}
\left(\frac{8m^3M^2}{Hs_1^2}\right)^{s_1}
\exp\left(-\frac{2\pi m^3M^2}{Hs_1^2}\right) .
\end{equation}
For specific $m$ and $H$ this can be integrated numerically in order
to obtain the change of the mass as a function of time.

\begin{figure}[!ht]
    \begin{center}
    \includegraphics[width=8.6cm]{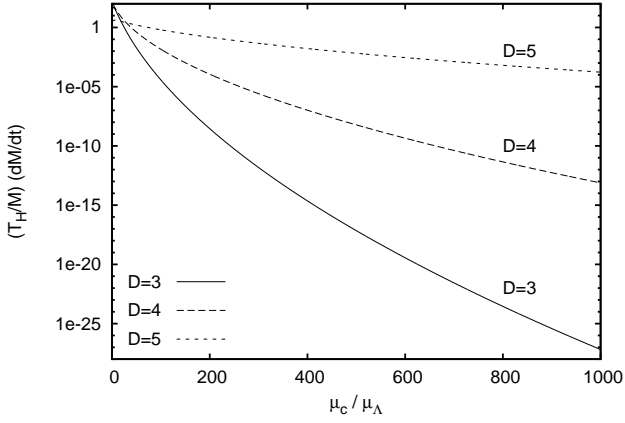}
    \end{center}
    \caption{The relative mass loss during a Hubble time period,
      $\frac{T_H}{M}\frac{dM}{dt}$, as a function of
      $\frac{\mu_c}{\mu_\Lambda}$ for $D=3$, $4$ and $5$ spatial
      dimensions.
      \label{figmassl}}
\end{figure}

Using the leading order expression \eqref{meps} for the total mass and
dividing by $M$ we can obtain the the relative mass loss rate. The
Hubble time $T_H=1/H$ describes the time-scale of the expansion of the
universe. The value of $T_H$ obtained from the present value of the
cosmological constant is given by \eqref{hubbletime} in Appendix
\ref{sec:cosmconst}. Dividing by both $M$ and $H$ we obtain the
relative mass loss extrapolated for a period corresponding to the
Hubble time. For $D>3$,
\begin{equation}
\frac{T_H}{M}\frac{dM}{dt}=-\frac{S_t^2}{(D-2)s_1h}
\exp\left(-\frac{\pi}{2h}\right) , \label{relmassl45}
\end{equation}
and for $D=3$,
\begin{equation}
\frac{T_H}{M}\frac{dM}{dt}=-\frac{S_t^2}{s_1h}
\left(\frac{2}{h}\right)^{s_1}
\exp\left(-\frac{\pi}{2h}\right)\,, \label{relmassl3}
\end{equation}
where the values of the constants $s_1$, $S_t$ are given in Table \ref{c1table}.
This result is independent on the units used to measure the time and
mass, and it depends on $H$, $m$ and $\varepsilon$ only through the
rescaled cosmological constant $h=H/(m\varepsilon^2)$. According to
\eqref{mumulambda} the square of the parameter $h$ is proportional to
the ratio of the central density $\mu_c$ to the effective energy
density corresponding to the cosmological constant $\mu_\Lambda$. On
fig.  \ref{figmassl} and on Table \ref{mltable} we give
$\frac{T_H}{M}\frac{dM}{dt}$ as a function of the density ratio
$\frac{\mu_c}{\mu_\Lambda}$.
\begin{table}[htbp]
\begin{tabular}{|c|c|c|c|c|c|c|}
\hline
  & \multicolumn{2}{|c|}{$D=3$}  & \multicolumn{2}{|c|}{$D=4$}
     & \multicolumn{2}{|c|}{$D=5$} \\ \cline{2-7}
  $\frac{T_H}{M}\frac{dM}{dt}$ & $h$ & $\frac{\mu_c}{\mu_\Lambda}$
    & $h$ & $\frac{\mu_c}{\mu_\Lambda}$ & $h$ & $\frac{\mu_c}{\mu_\Lambda}$\\
  \hline
  $1$ & $0.11807$ & $24.95$ & $0.20053$ & $39.00$
    & $0.42136$ & $73.81$\\
  $0.1$ & $0.09460$ & $38.87$ & $0.15073$ & $69.03$
    & $0.23793$ & $231.5$\\
  $10^{-2}$ & $0.07985$ & $54.55$ & $0.12139$ & $106.4$
    & $0.16998$ & $453.5$\\
  $10^{-3}$ & $0.06951$ & $71.99$ & $0.10188$ & $151.1$
    & $0.13327$ & $737.8$\\
  $10^{-4}$ & $0.06176$ & $91.19$ & $0.08791$ & $202.9$
    & $0.10999$ & $1083$\\
  $10^{-5}$ & $0.05569$ & $112.1$ & $0.07739$ & $261.9$
    & $0.09382$ & $1489$\\
  \hline
\end{tabular}
\caption{The rescaled cosmological constant $h$ and the density ratio
$\frac{\mu_c}{\mu_\Lambda}$ belonging to various relative mass loss
during a Hubble time period. \label{mltable}}
\end{table}

\section{Mass loss rate of oscillatons in the $\Lambda=0$ case}

\begin{figure}[!ht]
    \begin{center}
    \includegraphics[width=8.6cm]{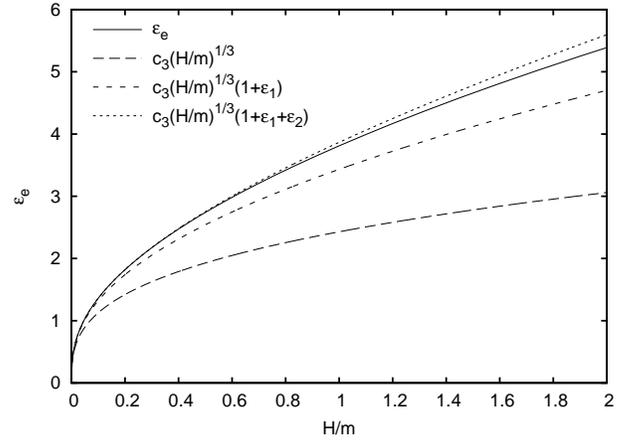}
    \end{center}
    \caption{The values of $\varepsilon_e$, for which the oscillaton mass
     loss rates \eqref{masslossd3} and \eqref{masslossflat} are equal are
     plotted as a function of the ratio of the Hubble constant $H$ and
     the scalar field mass $m$ (solid line).
     Above the line the asymptotically flat
     result \eqref{masslossflat} is dominant, while below
     \eqref{masslossd3} induced by the cosmological constant gives a
     larger contribution. The first three orders of the analytical
     approximation given by \eqref{epsilone} are also shown.
  \label{figcomp}}
\end{figure}

If the cosmological constant is zero and the space-time is
asymptotically flat then the metric of boson stars is static and their
mass is independent of time. In contrast, oscillatons are very slowly
losing mass by scalar field radiation even in the $\Lambda=0$ case
\cite{Page}. This mass loss is nonperturbatively small in the
amplitude parameter $\varepsilon$.  According to \cite{oscillaton},
for $D=3$ this mass loss rate depends only on the amplitude parameter,
and it is given by
\begin{equation}
\frac{\mathrm{d} M}{\mathrm{d} t}=
-\frac{c_1}{\varepsilon^{2}}
\exp\left(-\frac{c_2}{\varepsilon}\right) . \label{masslossflat}
\end{equation}
where the numerical values of the constants are
\begin{equation}
c_1=30.0 \,, \qquad c_2=22.4993 \,.
\end{equation}
For small $\varepsilon$ this is exponentially small, while the mass
loss induced by the cosmological constant, given by
\eqref{masslossd3}, tends to zero only polynomially when
$\varepsilon\to0$. This implies that if $\Lambda\not=0$ then for small
$\varepsilon$ the mass loss \eqref{masslossd3} originating from the
cosmological constant dominates, while for large $\varepsilon$ the
expression \eqref{masslossflat} can be applied. Apart from the
amplitude parameter $\varepsilon$ the mass loss \eqref{masslossd3}
depends only on the ratio $H/m$. It can be checked numerically that
for a concrete choice of $H/m$ there is only one $\varepsilon$ value
where the two type of mass loss rates are equal, which we denote by
$\varepsilon_e$.
For low $H/m$ values the function $\varepsilon_e$ can be expanded as
\begin{equation}
\varepsilon_e=c_3\left(\frac{H}{m}\right)^{1/3}
\left(1+\varepsilon_1+\varepsilon_2+\ldots\right) \,, \label{epsilone}
\end{equation}
with
\begin{equation}
\varepsilon_1=\frac{R}{3c_3^{2}}\left(\frac{H}{m}\right)^{1/3}  , \quad
\varepsilon_2=\frac{c_4R}{9c_3^{4}}\left(\frac{H}{m}\right)^{2/3}  ,
\end{equation}
\begin{equation}
R=\frac{2}{\pi}\ln\left(\frac{2^{s_1-1}S_t^2}{c_1}\right)
+c_4\ln c_3
+\left(\frac{c_4}{3}-\frac{2s_1}{\pi}\right)
\ln\left(\frac{H}{m}\right) ,
\end{equation}
where the newly introduced two constants are
\begin{equation}
c_3=\left(\frac{2c_2}{\pi}\right)^{1/3} \ , \quad
c_4=\frac{2}{\pi}\left(2s_1+5\right) \ .
\end{equation}
On Fig. \ref{figcomp} we plot $\varepsilon_e$ as a
function of $H/m$. To illustrate how well the above analytic approximation works
we plot for relatively large values of $H/m$, although the solutions
for $\epsilon>\epsilon_{\mathrm{max}}\approx0.5$ are unstable.
The quantity $H/m$ is originally assumed to be in Planck units, but it
has the same simple form in natural units if the Hubble constant is
expressed in electron volts.

\section{Possible effects in the current universe}

From the definition \eqref{caphlowh} of the rescaled cosmological
constant we can express the square of the small-amplitude parameter
$\varepsilon$. The amplitude of the scalar field is proportional to
$\varepsilon^2$ to leading order. Measuring the cosmological constant
in $1/s$ units and the scalar field mass in $eV/c^2$ units,
\eqref{caphlowh} can be written as
\begin{equation}
\varepsilon^2=\frac{6.58\times 10^{-16}}{h}\,Hs\,\frac{eV}{mc^2} \,.
\label{epstoh}
\end{equation}
According to \eqref{relmassl45} and \eqref{relmassl3}, the part of the
mass lost during a Hubble time period only depends on the value of the
rescaled Hubble constant $h$. For example, we can look for
configurations where this mass loss is one percent.  According to
Table \ref{mltable}, for the physically interesting $3+1$ dimensional
case this corresponds to the value of $h_{0.01}=0.07985$. Using the
present value of the Hubble constant, which is given by
\eqref{hubtpres},
\begin{equation}
\varepsilon^2_{0.01}=1.60\times 10^{-32}\frac{eV}{mc^2} \,. \label{epscur}
\end{equation}

Oscillatons or boson stars are generally stable if the amplitude
parameter $\varepsilon$, which for larger amplitudes can be defined
from the frequency $\omega$ by $\varepsilon=\sqrt{1-\omega^2/m^2},$ is
smaller than $\varepsilon_{\textrm{max}}\approx0.5$. From
\eqref{epscur} it is apparent that significant mass loss for the
larger amplitude configurations with
$\varepsilon\approx\varepsilon_{\textrm{max}}$ can happen only for
extremely small scalar field masses. Very small $\varepsilon$ values
are not excluded, but since for $D=3$, according to \eqref{mass3d},
the total mass of the oscillaton or boson star is proportional to
$\varepsilon$, these configurations tend to have extremely small
masses. Similarly, since by \eqref{eq:size} the radius is inversely
proportional to $\varepsilon$, small $\varepsilon$ values belong to
spatially extended configurations. It is obvious that the influence of
the cosmological constant is larger for large radius objects. In Table
\ref{curtabl} we list the values of $\varepsilon$, the total mass and
the radius for various choices of the scalar field mass $m$, assuming
that the mass-loss extrapolated for a Hubble time period is $1\%$.
\begin{table}[htbp]
\begin{tabular}{|c|l|l|l|}
\hline
 $mc^2/eV$  & $\quad\varepsilon_{0.01}$ & $\quad M/kg$
   & $\quad r_{95}/\mathrm{m}$ \\
  \hline
  $10^{10}$&$1.27\times10^{-21}$&$5.90\times10^{-11}$&$6.95\times10^{4}$\\
  $10^5$  & $4.01\times10^{-19}$&$1.87\times10^{-3}$&$2.20\times10^{7}$\\
  $1$     & $1.27\times10^{-16}$&$5.90\times10^{4}$&$6.95\times10^{9}$\\
  $10^{-5}$&$4.01\times10^{-14}$&$1.87\times10^{12}$&$2.20\times10^{12}$\\
  $10^{-10}$&$1.27\times10^{-11}$& $5.90\times10^{19}$&$6.95\times10^{14}$\\
  $10^{-15}$&$4.01\times10^{-9}$&$1.87\times10^{27}$&$2.20\times10^{17}$\\
  $10^{-20}$&$1.27\times10^{-6}$& $5.90\times10^{34}$&$6.95\times10^{19}$\\
  $10^{-25}$&$4.01\times10^{-4}$&$1.87\times10^{42}$&$2.20\times10^{22}$\\
  $10^{-30}$&$1.27\times10^{-1}$& $5.90\times10^{49}$&$6.95\times10^{24}$\\
  \hline
\end{tabular}
\caption{Oscillaton or boson star configurations for which $1\%$ of the
  mass is lost during Hubble time, assuming the present value of the
  cosmological constant. For each choice of the scalar field mass
  $m$ the amplitude parameter $\varepsilon$, the total mass $M$
  and the radius containing $95\%$ of the mass is given. \label{curtabl}}
\end{table}
Keeping the same scalar mass, for larger $\varepsilon$, and
consequently, for larger total masses and smaller radiuses, the mass
loss rate is smaller than $1\%$. For all configurations listed in
Table \ref{curtabl} the $\Lambda=0$ oscillaton mass loss rate given
by \eqref{masslossflat} is negligible, many orders of magnitude
smaller than $1\%$.

To show some concrete examples, we examine three specific choices for
the scalar field mass. Firstly, for an axion with $m=10^{-5}eV$ the
mass of the oscillaton for which $\frac{T_H}{M}\frac{dM}{dt}=0.01$ is
about that of a very small asteroid, and its radius is about $15$
astronomical units.  Secondly, if $m=9.55\times10^{-18}eV/c^2$, then
the total mass is about a solar mass, $M_\odot=2.00\times10^{30}kg$,
which is distributed in a region of radius of $238$
light-years. Thirdly, for $m=10^{-25}$ the corresponding mass is about
the mass of the Milky Way (including dark matter), and the radius is
$2\times10^6$ light-years, which is about four times the radius of the
stellar disk. According to Table \ref{mltable} and \eqref{cosmdens},
for all the above configurations the central density is
$\mu_c=54.55\mu_\Lambda=3.7\times10^{-22}kg/\mathrm{m}^3$.

The maximal amplitude boson stars or oscillatons, with
$\varepsilon\approx\varepsilon_{\textrm{max}}$ generally have much
bigger masses and smaller sizes than the states with $1\%$ mass loss
discussed in the previous paragraphs. For example, the
$\varepsilon=0.5$ oscillaton for the axion case has a mass
$M=2.3\times10^{25}kg$ (about four Earth masses) and radius
$r_{95}=18\mathrm{cm}$. Only considering the mass loss
\eqref{masslossd3} induced by the cosmological constant, the maximal
amplitude boson stars or oscillatons have extremely long lifetime. For
the axion case, using \eqref{epstoh}, we get $h=5.1\times10^{-28}$,
and the relative mass loss during a Hubble time period is of the order
$10^{-10^{27}}$, certainly negligible. For boson stars formed by
complex fields this is the only mechanism to lose mass, consequently,
close to maximal mass boson stars have practically constant
mass. However oscillatons lose mass even in the $\Lambda=0$ case by
\eqref{masslossflat} which gives
$\frac{T_H}{M}\frac{dM}{dt}=-3.1\times10^{10}$ for $\varepsilon=0.5$,
showing that this state is unstable on the cosmological
time-scale. However, the mass loss decreases exponentially with
$\varepsilon$. An initially maximal mass oscillaton created in the
early universe loses mass relatively quickly, and its amplitude
parameter decreases to about $\varepsilon=0.31$ during a Hubble time
period, where according to \eqref{masslossflat} the relative mass loss
rate is $\frac{T_H}{M}\frac{dM}{dt}=-0.14$. For more details and
different scalar field masses see Table VIII of \cite{oscillaton}.

\section{Inflationary era}

The energy density of the inflaton field during the inflationary epoch
can be estimated as $\mu_\Lambda=(10^{16}GeV)^4$ (see
e.g.~\cite{LythLiddle}). This expression is valid in natural units,
where $c=\hbar=1$ (but $G\not=1$) and the energy is measured in
electron volts. In ordinary units this corresponds to the mass density
$\mu_\Lambda=2.3\times10^{84}kg/\mathrm{m}^3$. Assuming a de Sitter
geometry, the Hubble constant is
\begin{equation}
H=\sqrt{\frac{8\pi G\mu_\Lambda}{3}}=3.6\times10^{37}\frac{1}{s}
=2.4\times10^{13}GeV  \,, \label{hubinf}
\end{equation}
in ordinary and natural units. By \eqref{caphlambda} this belongs to a
cosmological constant
\begin{equation}
\Lambda=\frac{3H^2}{c^2}=4.3\times10^{58}\frac{1}{\mathrm{m}^2}
=1.1\times10^{-11} \,,
\end{equation}
in ordinary and Planck units, respectively.

If in addition to the inflaton field there is another scalar field
$\chi$ on this de Sitter background, and it is massive, then it is
likely to form localized oscillaton configurations (or boson stars in
case of a complex field). This second field may be, for example, the
waterfall field in a hybrid inflation theory\cite{linde1,linde2}. If
the mass parameter of the field $\chi$ is $m$, then we can apply
\eqref{epstoh} to relate the small-amplitude parameter $\varepsilon$
to the rescaled Hubble constant $h$. The parameter $h$ is the
essential parameter determining the mass loss rate of the object. For
example, if the mass loss extrapolated for a Hubble time period is
assumed to be $1\%$, then $h=h_{0.01}=0.07985$, and for the
inflationary era we obtain
\begin{equation}
\varepsilon^2_{0.01}=3.0\times10^{14}\frac{GeV}{mc^2} \,.
\end{equation}
Since oscillatons and boson stars are stable only if the amplitude
parameter satisfies
$\varepsilon<\varepsilon_{\textrm{max}}\approx0.5$, states with $1\%$
mass loss can only exist for $m>1.2\times10^{15}GeV/c^2$. For objects
formed from fields with $m<1.2\times10^{15}GeV/c^2$, necessarily
$\varepsilon<\varepsilon_{\textrm{max}}<\varepsilon_{0.01}$. Then from
\eqref{epstoh} it follows that $h>h_{0.01}$, which implies that the
mass loss during the Hubble time is always larger than
$1\%$. Oscillatons and boson stars formed from scalar fields with mass
$m<10^{15}GeV/c^2$ are very short living during the inflationary
era. Another way to see why this lower limit is necessary is to
calculate the central density $\mu_c$ by \eqref{massendens}, and
observe that for small scalar field mass $\mu_c<\mu_\Lambda$, and
hence long living oscillatons are not expected to exist.

For scalar field masses $m>1.2\times10^{15}GeV/c^2$ we list the
properties of some oscillatons with $1\%$ mass loss in Table
\ref{infltabl}.
\begin{table}[htbp]
\begin{tabular}{|c|c|c|c|c|}
\hline
 $\frac{mc^2}{GeV}$  & $\varepsilon_{0.01}$ & $\frac{M}{kg}$
  & $\frac{r_{95}}{\mathrm{m}}$
  & $\frac{M_{\mathrm{max}}}{M}=\frac{r_{95}}{r_{95}^{\mathrm{min}}}$ \\
 \hline
 $1.2\times10^{15}$&$0.50$& $1.9\times10^{-4}$&$1.5\times10^{-30}$&$1.0$\\
 $10^{16}$&$0.17$& $8.0\times10^{-6}$&$5.1\times10^{-31}$&$2.9$\\
 $10^{17}$&$0.054$& $2.5\times10^{-7}$&$1.6\times10^{-31}$&$9.2$\\
 $10^{18}$&$0.017$& $8.0\times10^{-9}$&$5.1\times10^{-32}$&$29$\\
 $10^{19}$&$0.0054$& $2.5\times10^{-10}$&$1.6\times10^{-32}$&$92$\\
 \hline
\end{tabular}
\caption{Oscillaton or boson star configurations for which $1\%$ of the
  mass is lost during a Hubble time period in the inflationary
  era. For each scalar field mass $m$ the amplitude parameter
  $\varepsilon$, the total mass $M$, and the radius containing $95\%$
  of the mass is listed. In the last column we give the relation to the
  $\varepsilon=\varepsilon_{\textrm{max}}$ state, which has maximal
  mass and minimal radius.
 \label{infltabl}}
\end{table}
All these states have the central density
$\mu_c=54.55\mu_\Lambda=(2.7\times10^{16}GeV)^4
=1.3\times10^{86}kg/\mathrm{m}^3$, and the $\Lambda=0$ mass loss is
much smaller than $1\%$ for them.

The mass loss rate induced by the cosmological constant decreases very
quickly when considering higher amplitude configurations. The behavior
of the $\Lambda=0$ oscillaton mass loss is just the opposite. The two
kind of relative mass loss extrapolated for a Hubble time period of
some maximal amplitude states belonging to
$\varepsilon=\varepsilon_{\textrm{max}}\approx0.5$ are given in Table
\ref{maxinfl}.
\begin{table}[htbp]
\begin{tabular}{|c|c|c|c|c|}
\hline
 $\frac{mc^2}{GeV}$  & $h$ & $\frac{\mu_c}{\mu_\Lambda}$
  & $\frac{T_H}{M}\frac{dM}{dt}$
  & $\frac{T_H}{M}\frac{dM}{dt}\Bigr|_{\Lambda=0}$ \\
 \hline
 $10^{15}$&$0.095$& $38$& $0.10$&$1.7\times10^{-16}$\\
 $2\times10^{15}$& $0.047$& $150$& $1.5\times10^{-7}$&$3.3\times10^{-16}$\\
 $3\times10^{15}$& $0.032$& $350$& $5.8\times10^{-14}$&$5.0\times10^{-16}$\\
 $4\times10^{15}$& $0.024$& $620$& $1.4\times10^{-20}$&$6.6\times10^{-16}$\\
 $5\times10^{15}$& $0.019$& $970$& $2.3\times10^{-27}$&$8.3\times10^{-16}$\\
 \hline
\end{tabular}
\caption{
Relative mass loss rates extrapolated for a Hubble time period of
maximal amplitude configurations during the inflationary era. For each
choice of scalar field mass $m$ the rescaled Hubble constant $h$, the
ratio $\mu_c/\mu_\Lambda$, the mass loss rate induced by the
cosmological constant, and the $\Lambda=0$ oscillaton mass loss rate
are listed. \label{maxinfl}}
\end{table}

\section{Reheating}

Oscillatons are likely to form after the end of inflation, in the
reheating era, when the inflaton field is oscillating around the
potential minimum. These oscillatons are likely to influence the
efficiency of how the inflaton's energy is transferred to other fields.
Since oscillatons are concentrated energy lumps, their influence on
the formation of inhomogeneities may also be important. At the
reheating stage the equation of state can be well approximated by a
pressureless fluid, and hence the effective cosmological constant can
be taken to be zero.  For the lifetime of oscillatons in this era one
can apply the considerations given in \cite{oscillaton} which are
valid in the $\Lambda=0$ asymptotically flat case. The mass of the
inflaton field when it is near the vacuum value can be estimated as
$m=10^{13}GeV/c^2$ (see e.g. \cite{mukhanov}). The maximal amplitude
stable oscillaton which can be formed by this field, when
$\varepsilon=\varepsilon_{\textrm{max}}\approx0.5$, has the total mass
$M=0.023kg$ and radius $r_{95}=1.8\times10^{-28}\mathrm{m}$. The mass
and radius of smaller amplitude oscillatons formed by this field can
be easily obtained by using the fact that the total mass is
proportional to $\varepsilon$ and the radius is proportional to
$1/\varepsilon$.

If the cosmological constant is zero, then a natural time scale is the
oscillation period, $T_\omega=2\pi/\omega$. According to
\eqref{omegaepsilon}, for small $\varepsilon$ we can replace $\omega$ by
$m$. Taking the total mass from \eqref{meps2}, the part of the mass
lost during an oscillation period is
\begin{equation}
\frac{T_\omega}{M}\frac{dM}{dt}=-\frac{4\pi c_1}{s_1\varepsilon^3}
\exp\left(-\frac{c_2}{\varepsilon}\right) .
\end{equation}
This expression is independent of the scalar field mass $m$ and on the
units that we use for measuring time. Even for the maximal amplitude
oscillaton with $\varepsilon=0.5$ this gives a very tiny value,
$\frac{T_\omega}{M}\frac{dM}{dt}=-2.5\times10^{-17}$. This shows that
oscillatons formed in the reheating period perform a large number of
oscillations before the mass loss by emitting classical scalar field
radiation becomes apparent. However, since their oscillation period is
extremely short, $T_\omega=4.1\times10^{-37}s$, they lose about half
of their mass in $8.3\times10^{-21}s$.  On the other hand, since this
mass loss depends exponentially on $\varepsilon$, even during a period
corresponding to the lifetime of the universe the amplitude parameter
does not decrease much below $\varepsilon=0.166$, where
$\frac{T_H}{M}\frac{dM}{dt}=-0.13$.

\section{Conclusions} \label{conclusion}

We have constructed spherically symmetric, spatially localized time dependent
solutions in a large class
of scalar theories coupled to Einstein's theory of gravitation and a positive
cosmological constant, $\Lambda$. Examples include boson star-type objects (for a complex field)
and oscillatons (for a real field), considered previously
for vanishing cosmological constant. A positive constant
has important qualitative effects on boson stars, in that due to the repulsion induced
by $\Lambda$ only radiating solutions exist, leading to a mass loss.
For sufficiently small values of $\Lambda$ this effect is negligible, so boson stars
with almost unchanged mass may be present in our Universe.
Even when $\Lambda=0$ oscillatons radiate slowly, and the consequent mass loss
is maximal for large amplitudes.
There is an additional mass loss
of oscillatons induced by a nonzero $\Lambda>0$, which
dominates for small amplitudes (corresponding to large sizes).
The mass loss of an oscillaton during a Hubble time period,
is completely determined by the ratio $\mu_c/\mu_\Lambda$, where
$\mu_\Lambda$ is the energy density due to $\Lambda$ and $\mu_c$ is
that of the oscillaton in its center. Our computations are valid
in spatial dimensions $2<D<6$, although the case $D=3$ has been given special attention.
An important cosmological application of our results is for ''waterfall''-type
oscillatons appearing in hybrid inflation models.

\appendix

\section{Connection formulae} \label{app:con}

In this appendix we derive the formulae \eqref{bpmap} connecting the
amplitudes in the $y<1$ and $y>1$ regions. We follow the method
described in \cite{griffiths}.  Since the expressions for $Z$ obtained
by the WKB method are singular at $y=1$, it is necessary to find a
solution of \eqref{onedsch} which is valid in a region around this
point. Let us introduce a new radial coordinate by
\begin{equation}
y=1+x \,.
\end{equation}
We approximate $p^2$ for small $h$ by
\begin{equation}
p^2\approx y^2-1=x(x+2)\approx 2x \,.
\end{equation}
Introducing a rescaled coordinate $z$ by
\begin{equation}
x=-z\frac{h^{2/3}}{2^{1/3}} \,,
\end{equation}
\eqref{onedsch} takes the form
\begin{equation}
\frac{d^2Z}{dz^2}-zZ=0 \,.
\end{equation}
The solution can be written in terms of Airy functions
\begin{equation}
Z=a\,\mathrm{Ai}\,z+b\,\mathrm{Bi}\,z \,.
\end{equation}
For $z\gg 0$ this has the asymptotic form
\begin{equation}
Z=\frac{1}{\sqrt{\pi}z^{1/4}}\left[
\frac{a}{2}\exp\left(-\frac{2}{3}z^{3/2}\right)
+b\exp\left(\frac{2}{3}z^{3/2}\right)
\right] , \label{ztpd3}
\end{equation}
and for $z\ll 0$
\begin{align}
Z&=\frac{1}{\sqrt{\pi}(-z)^{1/4}}\biggl\{
a\sin\left[\frac{2}{3}(-z)^{3/2}+\frac{\pi}{4}\right] \notag\\
&\qquad\qquad\qquad
+b\cos\left[\frac{2}{3}(-z)^{3/2}+\frac{\pi}{4}\right]
\biggr\} . \label{ztpd45}
\end{align}

We consider two matching regions surrounding $y=1$, where $|z|\gg 1$
and at the same time $|x|\ll 1$. Both of these can hold simultaneously
if $h$ is small enough. First we continue through the matching region
with $x<0$. Substituting $y=1+x$ into \eqref{p0int1} and expanding in
power series around $x=0$, we obtain
\begin{equation}
\frac{i}{h}\int_1^y p_0 dy=\frac{1}{3h}(-2x)^{\frac{3}{2}}
+\mathcal{O}\left(x^{\frac{5}{2}}\right) \,.
\end{equation}
This result can also be obtained by simply approximating $p_0$ by
$\sqrt{2x}$.  In this region the influence of the $s_1$ term is
subleading, so for $3\leq D\leq 5$ dimensions we obtain both from
\eqref{z45da} and \eqref{zs1d3}
\begin{equation}
Z=\frac{A_+}{(-2x)^{1/4}}\exp\left[\frac{1}{3h}(-2x)^{3/2}\right] .
\end{equation}
Comparing with \eqref{ztpd3}, we get
\begin{equation}
a=0 \ , \qquad
b=\frac{\sqrt{\pi}}{(2h)^{1/6}}A_+ \ . \label{abm1}
\end{equation}

Continuing into the second matching region with $x>0$, \eqref{ztpd45}
yields
\begin{align}
Z&=\frac{b}{2\sqrt{\pi}(-z)^{1/4}}\biggl\{
\exp\left(\frac{i\pi}{4}\right)
\exp\left[i\frac{2}{3}(-z)^{3/2}\right] \notag\\
&\qquad\qquad+\exp\left(-\frac{i\pi}{4}\right)
\exp\left[-i\frac{2}{3}(-z)^{3/2}\right]
\biggr\} . \label{zxg0}
\end{align}
For small positive $x$ \eqref{p0int2} gives
\begin{equation}
\frac{i}{h}\int_1^y p_0 dy=\frac{i}{3h}(2x)^{\frac{3}{2}}
+\mathcal{O}\left(x^{\frac{5}{2}}\right) \,.
\end{equation}
To leading order both \eqref{z45db} and \eqref{zs1d45} can be written
as
\begin{equation}
Z=\frac{B_\pm}{(2x)^{1/4}}
\exp\left[\pm\frac{i}{3h}(2x)^{3/2}\right] .
\end{equation}
Comparing with \eqref{zxg0} gives
\begin{equation}
B_\pm=\frac{b(2h)^{1/6}}{2\sqrt{\pi}}
\exp\left(\pm\frac{i\pi}{4}\right) .
\end{equation}
Substituting b from \eqref{abm1} yields the desired formulae
\eqref{bpmap}.

\section{Mass in spherically symmetric asymptotically de Sitter
  space-times} \label{appmass}

Since the scalar field tends to zero exponentially, at large distances
the metric approaches the static Schwarzschild-de Sitter metric.
Considering $D+1$ dimensional space-times, in Schwarzschild area
coordinates the Schwarzschild-de Sitter metric has the form
\begin{equation}
ds^2=-F(\bar r)dt^2+\frac{1}{F(\bar r)}d\bar r^2
+\bar r^2d\Omega_{D-1}^2 \,,
\end{equation}
where
\begin{equation}
F(\bar r)=1-\frac{r_0^{D-2}}{\bar r^{D-2}}-H^2\bar r^2
\end{equation}
and $r_0$ is a constant related to the mass $M$ by
\begin{equation}
M=\frac{(D-1)\pi^{\frac{D}{2}}}
{8\pi\Gamma\left(\frac{D}{2}\right)}
r_0^{D-2}\,.
\end{equation}

In general spherically symmetric space-times there is a naturally
defined radius function $\hat r$, defined in terms of the area of the
symmetry spheres. In Schwarzschild coordinates $\hat r=\bar r$, while
in the isotropic coordinates \eqref{eq:metrgen} $\hat r=r \sqrt{B}$.
In terms of this radius function one can define the Misner-Sharp
energy (or local mass) function $\hat m$ \cite{MisnerSharp,nakao},
which can be defined for arbitrary dimensions by
\begin{equation}
\hat m=\frac{(D-1)\pi^{\frac{D}{2}}}
{8\pi\Gamma\left(\frac{D}{2}\right)}
\hat r^{D-2}\left(1-g^{ab}\hat r_{,a}\hat r_{,b}-H^2\hat r^2\right)
\,. \label{eq:massfunc}
\end{equation}
For the Schwarzschild-de Sitter metric $\hat m=M$. At infinity the
mass function $\hat m$ agrees with the Abbott-Deser mass
\cite{abbott,nakao2}, which is a Killing energy defined in static
asymptotically de Sitter space-times.

The Misner-Sharp energy $\hat m$ also agrees with a conserved energy
that can be defined using the Kodama vector. The Kodama vector
\cite{Kodama,Hayward} is defined by
\begin{equation}
K^a=\epsilon^{ab}\hat r_{,b} \,,
\end{equation}
where $\epsilon_{ab}$ is the volume form in the $(t,r)$
plane. Choosing the orientation such that $\epsilon_{rt}=\sqrt{AB}$
makes $K^a$ future pointing, with nonvanishing components
\begin{equation}
K^t=\frac{\hat r_{,r}}{\sqrt{AB}} \  , \qquad
K^r=-\frac{\hat r_{,t}}{\sqrt{AB}} \,.
\end{equation}
It can be checked that, in general, the Kodama vector is divergence
free, $K^a_{\ ;a}=0$. Since contracting with the Einstein tensor,
$G^{ab}K_{a;b}=0$, the current
\begin{equation}
J_a=T_{ab}K^b
\end{equation}
is also divergence free, $J^a_{\ ;a}=0$, it defines a conserved
charge. Integrating on a constant $t$ hypersurface with a future
oriented unit normal vector $n^a$, the conserved charge is
\begin{align}
E&=\frac{2\pi^{\frac{D}{2}}}{\Gamma\left(\frac{D}{2}\right)}
\int_0^r\hat r^{D-1}\sqrt{B}\,n^aJ_a dr  \label{eq:er}\\
&=\frac{2\pi^{\frac{D}{2}}}{\Gamma\left(\frac{D}{2}\right)}
\int_0^r\frac{\hat r^{D-1}}{A}\left(T_{tt}\hat r_{,r}
-T_{tr}\hat r_{,t}\right) dr \,. \notag
\end{align}
It can be checked that the derivative of the mass function $\hat m$
can be expressed in terms of the current $J_a$,
\begin{equation}
\hat m_{,a}=-\frac{2\pi^{\frac{D}{2}}\hat r^{D-1}}
{\Gamma\left(\frac{D}{2}\right)}
\epsilon_{ab}J^b \,.
\end{equation}
For the radial derivative follows that
\begin{equation}
\hat m_{,r}=\frac{2\pi^{\frac{D}{2}}\hat r^{D-1}}
{\Gamma\left(\frac{D}{2}\right)A}
\left(T_{tt}\hat r_{,r}
-T_{tr}\hat r_{,t}\right) \,,
\end{equation}
which, comparing with \eqref{eq:er}, gives $E=\hat m$.

The time derivative of the mass function is
\begin{equation}
\hat m_{,t}=\frac{2\pi^{\frac{D}{2}}\hat r^{D-1}}
{\Gamma\left(\frac{D}{2}\right)B}
\left(T_{rt}\hat r_{,r}
-T_{rr}\hat r_{,t}\right) \,. \label{masstder}
\end{equation}
This equation describes the mass loss caused by the outward energy
current of the massive scalar field.

For large radii the function $\hat m$ tends to the total mass $M$ of
the oscillon or boson star. Substituting the small-amplitude
expansions \eqref{eq:aexp} and \eqref{eq:bexp} into the definition
\eqref{eq:massfunc} of the mass function $\hat m$, to leading order in
$\varepsilon$ we obtain
\begin{equation}
\hat m=-\varepsilon^{4-D}\frac{\pi^{\frac{D}{2}}(D-1)\rho^{D-1}}
{8\pi m^{D-2}\Gamma\left(\frac{D}{2}\right)}
\,\frac{dB_2}{d\rho}\,.
\end{equation}
We have seen in Subsections \ref{secbsexp} and \ref{secoscexp} that
for both the boson star and oscillaton case $B_2=A_2/(2-D)$ and
$A_2=-1-s$, giving
\begin{equation}
\hat m=-\frac{\varepsilon^{4-D}\pi^{\frac{D}{2}}(D-1)\rho^{D-1}}
{8\pi(D-2)m^{D-2}\Gamma\left(\frac{D}{2}\right)}
\,\frac{ds}{d\rho}   \,. \label{hatmfn}
\end{equation}
Since the asymptotic behavior of $s$ is given by
\eqref{sasympt},
\begin{equation}
M=\varepsilon^{4-D}\frac{(D-1)\pi^{\frac{D}{2}}}
{8\pi m^{D-2}\Gamma\left(\frac{D}{2}\right)}s_1 \,. \label{meps}
\end{equation}
For $3+1$ dimensional space-time
\begin{equation}
M_{(D=3)}=\varepsilon \frac{s_1 }{2m}  \,. \label{meps2}
\end{equation}
For small $h$ the numerical value of the constant $s_1$ can be
approximated by the $h=0$ value given in Table \ref{c1table}.
Measuring the $mc^2$ belonging to the scalar field in electron volts
and the mass of the boson star or oscillaton in kilograms,
\begin{equation}
M_{(D=3)}=4.657\times 10^{20}kg\frac{eV}{mc^2}\varepsilon \,. \label{mass3d}
\end{equation}

\section{Proper mass and core radius}

The proper mass inside a sphere of radius $r$ is defined by the $D$
dimensional volume integral of the energy density $\mu$,
\begin{equation}
M_p(r)=\frac{2\pi^{\frac{D}{2}}}{\Gamma\left(\frac{D}{2}\right)}
\int_0^r dr\mu B^{D/2}r^{D-1} \,. \label{eq:massint}
\end{equation}
Substituting the leading order expression \eqref{massenlead} for
$\mu$, and the leading order value $B=1$ from \eqref{eq:bexp}, using
the rescaled radial coordinate $\rho=\varepsilon m r$,
\begin{equation}
M_p(\rho)=\frac{\varepsilon^{4-D}\pi^{\frac{D}{2}}(D-1)}
{8\pi(D-2)m^{D-2}\Gamma\left(\frac{D}{2}\right)}
\int_0^\rho d\rho\rho^{D-1}S^2 \,. \label{eq:massint2}
\end{equation}
The integral can be performed using the SN equation \eqref{seq},
\begin{equation}
M_p(\rho)=-\frac{\varepsilon^{4-D}\pi^{\frac{D}{2}}(D-1)\rho^{D-1}}
{8\pi(D-2)m^{D-2}\Gamma\left(\frac{D}{2}\right)}
\,\frac{ds}{d\rho} \,. \label{eq:massint3}
\end{equation}
To leading order in $\varepsilon$ the calculated proper mass $M_p$
agrees with the mass function $\hat m$ given by
\eqref{hatmfn}. However at higher order the proper mass is expected to
be larger by an amount corresponding to the binding energy.

Oscillatons and boson stars do not have a definite outer surface. A
natural definition for their size is to take the radius $r_n$ inside
which $n$ percentage of the mass-energy can be found. It is usual to
take, for example, $n=95$ or $99.9$. The mass-energy inside a given
radius $r$ can be defined either by the integral \eqref{eq:massint} or
by taking the local mass function $\hat m$ in \eqref{eq:massfunc}. To
leading order in $\varepsilon$ both definitions give
\eqref{eq:massint3}.  The rescaled radius $\rho_{n}$ can be defined by
\begin{equation}
\frac{M_p(\rho_{n})}{M_p(\infty)}=\frac{n}{100} \,.
\end{equation}
The numerical values of $\rho_{n}$ for various $n$ in $D=3,4,5$
dimensions are listed in Table \ref{tablerhon}.
\begin{table}[htbp]
\begin{tabular}{|c|c|c|c|}
\hline
  & $D=3$  & $D=4$  & $D=5$ \\
\hline
$\rho_{50}$   &  $2.240$ &   $1.778$ &  $1.317$ \\
$\rho_{90}$   &  $3.900$ &   $3.013$ &  $2.284$ \\
$\rho_{95}$   &  $4.471$ &   $3.455$ &  $2.652$ \\
$\rho_{99}$   &  $5.675$ &   $4.410$ &  $3.478$ \\
$\rho_{99.9}$ &  $7.239$ &   $5.692$ &  $4.634$ \\
\hline
\end{tabular}
\caption{\label{tablerhon}
  The radius inside which given percentage of the mass-energy is
  contained for various spatial dimensions.
}
\end{table}
The physical radius is
\begin{equation}
r_{n}=\frac{\rho_{n}}{\varepsilon m} \,.
\end{equation}
In ordinary units, measuring the scalar field $mc^2$ in electron volts
and $r_n$ in meters (Roman $\mathrm{m}$),
\begin{equation}
r_{n}=1.97\times 10^{-7}\mathrm{m}\,\frac{\rho_{n}}{\varepsilon}
\,\frac{eV}{mc^2} \,. \label{eq:size}
\end{equation}

We note that for complex field boson stars the current
\begin{equation}
j^a=\frac{i}{2}g^{ab}\left(\Phi^*\Phi_{,b}-\Phi^*_{,b}\Phi\right)
\end{equation}
can be used to define a conserved quantity. The conserved particle
number $N$ can be defined by integrating the time component $j^0$ for
a spacelike slice. To leading order in $\varepsilon$ the function
$j^0$ is proportional to the density $\mu$ given in
\eqref{massenlead}. Hence, to this order, the use of the particle
number instead of the mass-energy yields the same result for the
radius of boson stars.

\section{Present value of the cosmological constant}
\label{sec:cosmconst}

For the present observational value of the Hubble constant we take
\begin{equation}
H_0=70.4\frac{\mathrm{km}}{sMpc}=2.28\times 10^{-18}\frac{1}{s}
\end{equation}
according to the combined 7-year WMAP data \cite{komatsu}. To this
corresponds a critical density
\begin{equation}
\mu_{cr}=\frac{3H_0^2}{8\pi G}=9.31\times 10^{-24}\frac{kg}
{\mathrm{m}^3} \,.
\end{equation}
The fraction of this belonging to the dark energy provided by the
cosmological constant is given by $\Omega_\Lambda=0.728$, according to
WMAP results \cite{komatsu}. Consequently, the energy density
belonging to the cosmological constant is
\begin{equation}
\mu_\Lambda=\mu_{cr}\Omega_\Lambda
=6.78\times 10^{-24}\frac{kg}{\mathrm{m}^3} \,.  \label{cosmdens}
\end{equation}
According to \eqref{mulambda}, this corresponds to a cosmological
constant of
\begin{equation}
\Lambda=\frac{8\pi G\mu_\Lambda}{c^2}
=1.26\times 10^{-52}\frac{1}{\mathrm{m}^2} \,.
\end{equation}
In Planck units, since the Planck length is
$l_P=1.616252\times10^{-35}\mathrm{m}$,
\begin{equation}
\Lambda=3.30\times 10^{-122} \,.
\end{equation}
Since we are interested in the effect of the cosmological constant, we
neglect the ordinary and dark matter content of the universe. Then the
space-time can be described by a de Sitter metric with Hubble constant
$H$ given by \eqref{caphlambda}. In Planck, ordinary and natural units
\begin{equation}
H=1.05\times 10^{-61}=1.95\times 10^{-18}\frac{1}{s}
=1.28\times10^{-33}eV \,. \label{hubtpres}
\end{equation}
The Hubble time is then
\begin{equation}
T_H=\frac{1}{H}=5.14\times 10^{17}s=16.3\mathrm{Gyr} \,.
\label{hubbletime}
\end{equation}
Even though the ordinary and dark matter content have been neglected,
this value is still not too far from the age of the universe, which
according to WMAP data is $13.75\mathrm{Gyr}$ assuming the
$\mathrm{\Lambda CDM}$ model \cite{komatsu}.

\begin{acknowledgments}

This research has been supported by OTKA Grants No. K61636,
NI68228, and by the U.S. Department of Energy (D.O.E.) under
cooperative research agreement DE-FG 0205ER41360.
\end{acknowledgments}

\end{document}